\documentclass[twocolumn,amsmath,floatfix]{revtex4}
\usepackage{latexsym}
\usepackage{amssymb}
\usepackage{pdfpages}
\usepackage{graphics}
\usepackage{tikz}
\usepackage{tikz-feynman}
\usetikzlibrary{decorations.pathmorphing,decorations.markings}
\usetikzlibrary{shapes,decorations.pathmorphing}
\tikzfeynmanset{compat=1.1.0}

\begin{document}
\newcommand{\ja}{Jakuba\ss a-Amundsen }
\newcommand{\bfx}{\mbox{\boldmath $x$}}
\newcommand{\bfq}{\mbox{\boldmath $q$}}
\newcommand{\bfnabla}{\mbox{\boldmath $\nabla$}}
\newcommand{\bfalpha}{\mbox{\boldmath $\alpha$}}
\newcommand{\bfsigma}{\mbox{\boldmath $\sigma$}}
\newcommand{\bfeps}{\mbox{\boldmath $\epsilon$}}
\newcommand{\bfA}{\mbox{\boldmath $A$}}
\newcommand{\bfP}{\mbox{\boldmath $P$}}
\newcommand{\bfe}{\mbox{\boldmath $e$}}
\newcommand{\bfn}{\mbox{\boldmath $n$}}
\newcommand{\bfW}{{\mbox{\boldmath $W$}_{\!\!rad}}}
\newcommand{\bfM}{\mbox{\boldmath $M$}}
\newcommand{\bfI}{\mbox{\boldmath $I$}}
\newcommand{\bfJ}{\mbox{\boldmath $J$}}
\newcommand{\bfQ}{\mbox{\boldmath $Q$}}
\newcommand{\bfY}{\mbox{\boldmath $Y$}}
\newcommand{\bfp}{\mbox{\boldmath $p$}}
\newcommand{\bfk}{\mbox{\boldmath $k$}}
\newcommand{\bfks}{\mbox{{\scriptsize \boldmath $k$}}}
\newcommand{\bfqs}{\mbox{{\scriptsize \boldmath $q$}}}
\newcommand{\bfxs}{\mbox{{\scriptsize \boldmath $x$}}}
\newcommand{\bfalphas}{\mbox{{\scriptsize \boldmath $\alpha$}}}
\newcommand{\bfs}{\mbox{\boldmath $s$}_0}
\newcommand{\bfv}{\mbox{\boldmath $v$}}
\newcommand{\bfw}{\mbox{\boldmath $w$}}
\newcommand{\bfb}{\mbox{\boldmath $b$}}
\newcommand{\bfxi}{\mbox{\boldmath $\xi$}}
\newcommand{\bfzeta}{\mbox{\boldmath $\zeta$}}
\newcommand{\bfr}{\mbox{\boldmath $r$}}
\newcommand{\bfrs}{\mbox{{\scriptsize \boldmath $r$}}}
\newcommand{\bfps}{\mbox{{\scriptsize \boldmath $p$}}}
\renewcommand{\theequation}{\arabic{section}.\arabic{equation}}
\renewcommand{\thesection}{\arabic{section}}
\renewcommand{\thesubsection}{\arabic{section}.\arabic{subsection}}

\title{\Large\bf Radiative corrections to the parity-violating spin asymmetry}

\author{D.H.Jakubassa-Amundsen$^1$ and X.Roca-Maza$^{2,3,4,5}$}

\email{dj@math.lmu.de}

\email{xavier.roca.maza@fqa.ub.es}

\affiliation{$^1$Mathematics Institute, University of Munich, Theresienstrasse 39, 80333 Munich, Germany}

\affiliation{$^2$Departament de F\'isica Qu\`antica i Astrof\'isica, Mart\'i i Franqu\'es, 1, 08028 Barcelona, Spain}

\affiliation{$^3$Dipartimento di Fisica ``Aldo Pontremoli'', Universit\`a degli Studi di Milano, 20133 Milano, Italy}

\affiliation{$^4$INFN, Sezione di Milano, 20133 Milano, Italy}

\affiliation{$^5$Institut de Ci\`encies del Cosmos, Universitat de Barcelona, Mart\'i i Franqu\'es, 1, 08028 Barcelona, Spain}




\vspace{1cm}

\begin{abstract}

The parity-violating spin asymmetry  $A_{\rm pv}$ for elastic electron scattering from spin-zero nuclei, together with its QED corrections, is evaluated
nonperturbatively within the phase-shift analysis.
Dispersion corrections, taking into account low-lying transient nuclear excitations, are estimated
with the help of the lowest-order $\gamma Z$ box diagrams.
Collision energies between $5-500$ MeV are considered, and results are provided for the $^{12}$C and $^{208}$Pb target nuclei.
In addition, we have evaluated $A_{\rm pv}$ at GeV energies and small scattering angles --relevant for the Pb Radius Experiment (PREx)-- revealing that the low-lying nuclear excited states give no measurable dispersive contribution.
However, they are sizeable at lower energies and backward angles.

\end{abstract}

\maketitle

\section{Introduction}
\label{intro}

High-precision measurements of the parity-violating spin asymmetry $A_{\rm pv}$ at GeV collision energies have become feasible during the recent years \cite{Ad21,Ad22,An22},
improving earlier experiments \cite{Ab12}.
Since $A_{\rm pv}$ depends strongly on the neutron ground-state distribution and on the Weinberg mixing angle, it serves to extract the target neutron skin \cite{Ho98,Ho01,Re21}, but also provides information on the weak neutral current couplings and the weak charge radius \cite{Do89,Ho12,Ca24}.
For this application of $A_{\rm pv}$ it is mandatory for theory to take into account any possible corrections as precisely as possible.
Here we consider the radiative corrections from vacuum polarization and the vertex plus self-energy correction, as well as from dispersion, based on the excitation of low-lying  nuclear states.
Dispersion estimates are usually obtained from the $\gamma Z$ box diagrams \cite{GH09,Go11} and are used to correct the weak charge $Q_{\rm w}$ which enters into the normalization of the weak potential $A_{\rm w}$ \cite{Ho98,Ko20}.
Also the universal radiative corrections, estimated in Born approximation, are included in the modification of $Q_{\rm w}$ \cite{GH09,Nav24}.
An exact calculation of dispersion involves the sum over all transient excited nuclear states.
Several approximations have been in use. The earliest one is the closure approximation where all nuclear excitations are taken at
the same energy, which lies preferrably in the giant dipole resonance region \cite{FR74}.
For lower collision energies it is more adequate to consider explicitly the most collective excited states with small angular momentum and extending up to about 30 MeV \cite{Jaku22}.
A different approach, applicable for higher collision energies above the pion production threshold (135 MeV),
relies on the relation between dispersion, respectively the forward nuclear Compton amplitude, and the experimental photoproduction cross section \cite{AM04,GH08,Ko21}.
An application to the $\gamma Z$ box diagram can be found in \cite{Go11,Ha14}. Alternatively, the relation between dispersion and the structure functions from parity-violating deep inelastic scattering can be employed \cite{GH09}.

The present work is aimed at an interpretation of future precision experiments at the MESA facility on the parity-violating asymmetry for low-energy electron scattering \cite{Au11}.
Hence the QED effects are treated nonperturbatively by expressing them in terms of potentials \cite{Ue35,Jaku24,RJ26} which are included in the Dirac equation for the electronic scattering states.
Moreover, in the $\gamma Z$ box diagrams, the low-lying nuclear excitations are considered explicitly in terms of their transition densities as obtained from nuclear Energy Density Functionals \cite{Sch19,Co13,CM21}.
By inserting the dispersion correction  directly into  the differential scattering cross section,  one is independent of the widely used Born representation of $A_{\rm pv}$.

The paper is organized as follows:
Sec.~\ref{theory} describes the theory for $A_{\rm pv}$ and its radiative corrections. Results for the $^{12}$C and $^{208}$Pb target nuclei are presented in Sec.~\ref{results}. Concluding remarks follow in Sec.~\ref{conclusion}.
Atomic units ($\hbar=m_e=e=1)$ are used unless indicated otherwise.

\section{Theory}
\label{theory}

In this section the theory for $A_{\rm pv}$ is shortly recapitulated.  Subsequently, the treatment of  the QED corrections is presented.
Finally dispersion is considered to lowest order.

\subsection{Parity-violating spin asymmetry}

For collision energies well above 10 MeV, the electron mass $m_e$ can usually be neglected. Hence the electrons polarized longitudinally, i.e. parallel (+) or antiparallel ($-$) to
their direction of motion, are described by the helicity eigenstates $\psi_+$, respectively $\psi_-$, which are solutions to the Dirac equation
\begin{equation}\label{2.1}
[-ic\bfalpha \bfnabla + V(r) \pm A(r)]\;\psi_\pm(\bfr)\,=\,E\;\psi_\pm(\bfr),
\end{equation}
where, to start with, $V(r)=V_T(r)$ is taken as the Coulombic nuclear target potential and $A(r)=A_{\rm w}(r)$ as  the weak potential which is
related to the weak density $\varrho_{\rm w}(r)$ by means of \cite{Ho98}
\begin{equation}\label{2.2}
A_{\rm w}(r)\,=\,\frac{G_Fc}{2\sqrt{2}}\;\varrho_{\rm w}(r),
\end{equation}
where $G_F=1.16638 \times 10^{-5} $ GeV$^{-2}$, equivalent to $G_Fc=6.2283 \times 10^{-5}$ fm$^{2}$, is the Fermi coupling constant.

The weak density  is calculated from the convolution of the point-like proton ($\varrho_p$, normalized to the number of protons $Z$) and neutron ($\varrho_n$, normalized to the number $N$ of neutrons) nuclear density distributions with the proton ($G_p^{\rm w}$) and neutron ($G_n^{\rm w}$) weak form factors
as in \cite{Ho01,Roc11,Rei21},
\begin{equation}\label{3.6a}
\varrho_{\rm w}(\bfr) = \int d\bfr^\prime \left[G_p^{\rm w}(\bfr^\prime)\varrho_p(\bfr-\bfr^\prime) + G_n^{\rm w}(\bfr^\prime)\varrho_n(\bfr-\bfr^\prime)\right] \ . 
\end{equation}
The latter  ones can be written in terms of the electromagnetic form factors ($G_{n,p}^{\rm em}$) and the weak charges of the neutron ($Q_n^{\rm w}$) and the proton ($Q_p^{\rm w}$) in the usual way,  
\begin{eqnarray}
G_p^{\rm w} &=& Q_p^{\rm w} G_p^{\rm em}+ Q_n^{\rm w} G_n^{\rm em}\\
G_n^{\rm w} &=& Q_n^{\rm w} G_p^{\rm em}+ Q_p^{\rm w} G_n^{\rm em}
\end{eqnarray}
where at tree level in the Standard Model, with $\theta_{\rm w}$ the Weinberg mixing angle,
\begin{eqnarray}
Q_p^{\rm w} &=& 1-4\sin^2 \theta_{\rm w} \\
Q_n^{\rm w} &=& - 1.
\end{eqnarray}

The parity-violating (or 'weak') spin asymmetry is obtained from the relative cross section difference,
\begin{equation}\label{2.3}
A_{\rm pv}\;=\;\frac{d\sigma/d\Omega(+)-d\sigma/d\Omega(-)}{d\sigma/d\Omega(+)+ d\sigma/d\Omega(-)},
\end{equation}
where the differential scattering cross section for the helicity ($+$) and ($-$) states is given by
\begin{equation}\label{2.4}
\frac{d\sigma_{\rm coul}}{d\Omega}(\pm)\,=\,\frac{|\bfk_f|}{|\bfk_i|}\;\frac{1}{f_{\rm rec}} \sum_{\sigma_f}  \;\left| A_{fi}^{\rm coul}(\psi_\pm)\right|^2.
\end{equation}

We have attached the superscript 'coul' to the scattering amplitude $A_{fi}$ to indicate   that the potential $V(r)$ used in the Dirac equation is
solely the Coulombic nuclear potential.
The momenta of the incoming and scattered electron are denoted by $k_i$ and $k_f$, respectively, 
$\sigma_f$ is the final spin polarization and $f_{\rm rec}$ is  a recoil factor close to unity \cite{DS84}.

\subsection{QED corrections}

The scattering amplitude $A_{fi}^{\rm QED}$, which includes the QED effects, is obtained
from the wavefunctions $\psi_\pm$ which are solutions to (\ref{2.1}) in the modified potentials,
$$
V(r)\,=\,V_T(r) + V_{\rm vac}(r) +V_{\rm vs}(r).
$$
\begin{equation}\label{2.5}
A(r)\,=\,A_{\rm w}(r)\,+\,A_{\rm vs}^{\rm ax}(r).
\end{equation}

The potential $V_{\rm vac}$ for vacuum polarization is taken to be the Uehling potential as parametrized in \cite{FR76}.
The potential $V_{\rm vs}$ for the vector vertex plus self-energy (vs) correction is defined in terms of the integral representation \cite{Jaku24},
\begin{equation}\label{2.6}
V_{\rm vs}(r)\,=\,-\,\frac{2Z}{\pi} \int_0^\infty d|\bfq|\;\frac{\sin(|\bfq|r)}{|\bfq|r}\;F_L(|\bfq|)\;F_1^{\rm vs}(-q^2),
\end{equation}
where 
 $q^2=(E_i-E_f)^2/c^2-\bfq^2$ with $\bfq=\bfk_i-\bfk_f$ is the squared 4-momentum transfer to the nucleus.
$E_i$ and $E_f$ are, respectively, the total energies of the incoming and scattered electron. $F_L$ is the nuclear ground-state charge form factor 
and $F_1^{\rm vs}$ is the electric form factor describing the vs correction in the first-order Born approximation \cite{T60,Va00}.

The potential $A_{\rm vs}^{\rm ax}$ for the axial-vector vs correction is in analogy to (\ref{2.6})
defined by means of the integral representation 
\cite{RJ26},
$$ A_{\rm vs}^{\rm ax}(r)\,=\, \frac{G_Fc}{4\sqrt{2} \pi^2} \int_0^\infty \bfq^2 d|\bfq|\,\frac{\sin(|\bfq|r)}{|\bfq|r}$$
\begin{equation}\label{2.11a}
\times \; F_{\rm w}(|\bfq|)\,F_1^{\rm w,vs}(-q^2),
\end{equation}
where $F_{\rm w}$ is the  form factor relating to the weak density and $F_1^{\rm w,vs}$ is the weak vs form factor describing the axial-vector vs correction in the first-order Born approximation.
Its large-$q$ limit is given by \cite{RH26}
$$F_1^{\rm w,vs}(-q^2)=\frac{1}{2\pi c}\left[ -\frac12 \ln^2(\frac{-q^2}{c^2}) +\frac{3}{2}\ln (\frac{-q^2}{c^2})+\frac{\pi^2}{6}-1\right],$$
\begin{equation}\label{2.11b}
\hspace{3cm} -q^2/c^2 \gg 1.
\end{equation}
In order to check the relevance of low momentum transfer in (\ref{2.11a}), it is assumed that the exact $q$-dependence is the same as for the electric vs form factor $F_1^{\rm vs}$.
Hence with
$$
F_0(-q^2)\,=\,\frac{\nu^2+1}{4\nu}\;(\ln \frac{\nu+1}{\nu-1})\;(\ln \frac{\nu^2-1}{4\nu^2})
$$
\begin{equation}\label{2.11c}
+\frac{2\nu^2+1}{2\nu}\ln\frac{\nu+1}{\nu-1}\,+\frac{\nu^2+1}{2\nu}\left\{ \mbox{Li}(\frac{\nu+1}{2\nu})-\mbox{Li}(\frac{\nu-1}{2\nu})\right\},
\end{equation}
where $\nu=\sqrt{1-4c^2/q^2}$ and Li$(x)=-\int_0^\infty dt \frac{\ln|1-t|}{t}$ is the Spence function, one has
$F_1^{\rm vs}(-q^2) ~ = ~ \frac{1}{2\pi c} ~ [F_0(-q^2)-2]$,
\begin{equation}\label{2.11d}
F_1^{\rm w,vs}(-q^2)\,=\,\frac{1}{2\pi c}\;[F_0(-q^2)-1].
\end{equation}
It turns out that, due to its decrease as $|\bfq|\to 0$, the small-$q$ part of the integrand in (\ref{2.11a}) can be neglected except for very low collision energies or near-forward scattering  angles (note that $-q^2/c^2 < 100$ corresponds to $|\bfq|<0.026$ fm$^{-1}$).

In concord with (\ref{2.4}), the differential cross section is given by
\begin{equation}\label{2.7}
\frac{d\sigma^{\rm QED}}{d\Omega}(\pm)\,=\,\frac{|\bfk_f|}{|\bfk_i|}\,\frac{1}{f_{\rm rec}}\,\sum_{\sigma_f} \left| A_{fi}^{\rm QED} (\psi_\pm)\right|^2.
\end{equation}

The corresponding parity-violating  spin asymmetry, termed $A_{\rm pv}^{\rm QED}$, is  defined by (\ref{2.3}),  with $d\sigma/d\Omega$ replaced by $d\sigma^{\rm QED}/d\Omega$. 

\subsection{Dispersion effects}

\begin{figure}
\begin{tikzpicture}
  \begin{feynman}
    \vertex (e1) at (0,2);
    \vertex (e2) at (0,-2);
    \vertex (v1) at (0,0.9) [dot]{};
    \vertex (v2) at (0,-0.9) [dot]{};

    \vertex (n1) at (2,2);
    \vertex (n2) at (2,-2);
    \vertex (v4) at (2,0.9) [dot]{};
    \vertex (v3) at (2,-0.9) [dot]{};

    \draw[fermion, thick, -] (e2) -- (v2);
    \draw[fermion, thick, -] (v2) -- (v1);
    \draw[fermion, thick, -] (v1) -- (e1);

    \draw[double, thick] (n2) -- (v3);
    \draw[double, thick] (v4) -- (n1);

    \draw[fill=gray!30] (2,0) ellipse (0.25 and 0.84);
    \node[right] at (2.3, 0) {\(N^*\)};

    \draw[thick, decorate, decoration={snake, segment length=6pt, amplitude=2pt}, postaction={decorate, draw=black, decoration={markings, mark=at position 1 with {\arrow[scale=1.5,fill=black]{latex}}}}]
        (v2) -- node[below right=1pt and -1pt] {\(\gamma\)} (v3);
    \draw[thick, decorate, decoration={snake, segment length=6pt, amplitude=2pt}, postaction={decorate, draw=black, decoration={markings, mark=at position 1 with {\arrow[scale=1.5,fill=black]{latex}}}}]
        (v1) -- node[above right=1pt and -1pt] {\(Z\)} (v4);

    \node[left] at (e1) {\(e\)};
    \node[left] at (e2) {\(e\)};
    \node[right] at (n1) {\(N\)};
    \node[right] at (n2) {\(N\)};

    \node at (0.9, 0.6) {\(q_2\)};
    \node at (0.9, -0.6) {\(q_1\)};
    \node at (-0.5, -1.6) {\(k_i\)};
    \node at (-0.5, 0) {\(p\)};
    \node at (-0.5, 1.6) {\(k_f\)};
    \node at (1, -2.5) {(a)};
    
  \end{feynman}
\end{tikzpicture}
\begin{tikzpicture}
  \begin{feynman}
    \vertex (e1) at (0,2);
    \vertex (e2) at (0,-2);
    \vertex (v1) at (0,0.9) [dot]{};
    \vertex (v2) at (0,-0.9) [dot]{};

    \vertex (n1) at (2,2);
    \vertex (n2) at (2,-2);
    \vertex (v4) at (2,0.9) [dot]{};
    \vertex (v3) at (2,-0.9) [dot]{};

    \draw[fermion, thick, -] (e2) -- (v2);
    \draw[fermion, thick, -] (v2) -- (v1);
    \draw[fermion, thick, -] (v1) -- (e1);

    \draw[double, thick] (n2) -- (v3);
    \draw[double, thick] (v4) -- (n1);

    \draw[fill=gray!30] (2,0) ellipse (0.25 and 0.84);
    \node[right] at (2.3, 0) {\(N^*\)};

    \draw[thick, decorate, decoration={snake, segment length=6pt, amplitude=2pt}, postaction={decorate, draw=black, decoration={markings, mark=at position 1 with {\arrow[scale=1.5,fill=black]{latex}}}}]
        (v2) -- node[below right=1pt and -1pt] {\(Z\)} (v3);
    \draw[thick, decorate, decoration={snake, segment length=6pt, amplitude=2pt}, postaction={decorate, draw=black, decoration={markings, mark=at position 1 with {\arrow[scale=1.5,fill=black]{latex}}}}]
        (v1) -- node[above right=1pt and -1pt] {\(\gamma\)} (v4);

    \node[left] at (e1) {\(e\)};
    \node[left] at (e2) {\(e\)};
    \node[right] at (n1) {\(N\)};
    \node[right] at (n2) {\(N\)};

    \node at (0.9, 0.6) {\(q_2\)};
    \node at (0.9, -0.6) {\(q_1\)};
    \node at (-0.5, -1.6) {\(k_i\)};
    \node at (-0.5, 0) {\(p\)};
    \node at (-0.5, 1.6) {\(k_f\)};
    \node at (1, -2.5) {(b)};
    
  \end{feynman}
\end{tikzpicture}

\caption{Feynman box diagrams for dispersion. Diagram (a) describes the electron (e) nucleus (N) coupling by a photon ($\gamma$) with momentum $q_1$ and subsequently by a $Z$ boson with momentum $q_2$. $N^*$ indicates the excitation of the nucleus $N$ while $k_i$ and $k_f$ are 
the momenta of the incoming and scattered electron.
In diagram (b), the order is reversed. \label{fig1}}
\end{figure}
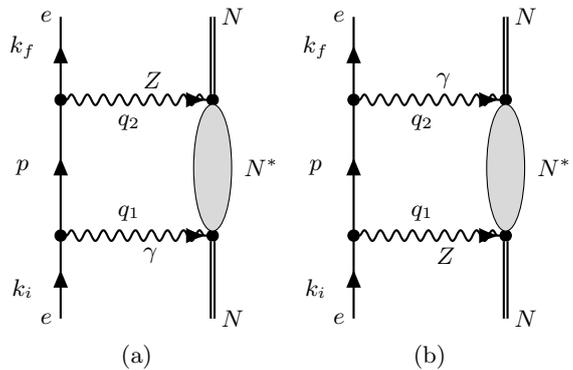

In lowest-order Born approximation, dispersion is calculated from the two Feynman box diagrams shown in Fig.~\ref{fig1}, involving one virtual photon and one $Z$ boson.
The contributions from the crossed-box diagrams are found to be small, and also the seagull diagrams can be disregarded according to \cite{FR74}.

The initial and final electronic helicity eigenstates are described by plane waves,
$$\psi_{i,\pm}^{(0)}\;=\;u_{k_i}^{(\pm)}$$
\begin{equation}\label{2.8}
\psi_{f,+}^{(0)}\;=\;\cos \frac{\theta}{2}\;u_{k_f}^{(+)}\;+\;\sin \frac{\theta}{2}\;u_{k_f}^{(-)}
\end{equation}
$$ \psi_{f,-}^{(0)}\;=\;-\;\sin \frac{\theta}{2}\;u_{k_f}^{(+)}\;+\;\cos \frac{\theta}{2}\;u_{k_f}^{(-)},$$
where $u_k^{(\pm)}$ are plane-wave 4-spinors (with the prefactor $\sqrt{(E+c^2)/(2E)}$, retaining the electron mass)  and $\theta$ is the scattering angle.
A coordinate system is chosen where the $z$-axis is along $\bfk_i$, while $\bfk_f$ lies in the $(x,z)$-plane.
The cross section with inclusion of the dispersion amplitude $A_{fi}^{\rm box}$, as calculated from the box diagrams of Fig.~\ref{fig1}, is to lowest order given by
$$\frac{d\sigma^{\rm box}}{d\Omega}(\pm)\;=\;\frac{|\bfk_f|}{|\bfk_i|}\;\frac{1}{f_{\rm rec}}\sum_{\sigma_f}  [ \;|A_{fi}^{\rm coul}(\psi_\pm)|^2 $$
\begin{equation}\label{2.9}
 +\;2\mbox{ Re } \{ A_{fi}^{\rm coul \ast}(\psi_\pm) \;A_{fi}^{\rm box}(\psi_\pm^{(0)}) \} ],
\end{equation}
$$A_{fi}^{\rm box}(\psi_\pm^{(0)})\;=\;2\sqrt{\frac{E_iE_f}{c^2}}\sum_{L,\omega_L} \left[ M_{fi}^{Z\gamma}(L,\omega_L,\pm)\,\right.$$
$$\left. +\,M_{fi}^{\gamma Z}(L,\omega_L, \pm)\right],$$
where $L$ is the angular momentum and $\omega_L$ the corresponding energy of the transient nuclear excited state.
The transition matrix element $M_{fi}^{Z\gamma}$ corresponds to the first and $M_{fi}^{\gamma Z}$ to the second Feynman diagram of Fig.~\ref{fig1}.
The evaluation of $M_{fi}^{Z\gamma}$ and $M_{fi}^{\gamma Z}$ proceeds as in the  case of the beam-normal spin asymmetry \cite{Jaku22}.
Details are given in the Appendices A and B.

The contribution of the axial  part of the $Z$-nucleus vertex  is 
 suppressed by the small electron's weak charge $Q_e^{\rm w}=-Q_p^{\rm w}\approx -0.05 $ \cite{Go11}.
Moreover, it represents magnetic transitions to excited nuclear states which are in general only relevant at the backmost  angles,
i.e. in
 a small portion of the phase space of the electron in its intermediate state.
Therefore this axial contribution is disregarded.

\section{Results}
\label{results}
\setcounter{equation}{0}
In this section we present estimates for the parity-violating spin asymmetry and its radiative corrections.
To this aim, the Dirac equation is solved with the help of the
Fortran code RADIAL \cite{Sal}.
The spin asymmetry modification $\Delta A_{\rm pv}$ is defined by
\begin{equation}\label{3.1}
\Delta A_{\rm pv}^{\rm QED/box}\,=\,A_{\rm pv}^{\rm QED/box}\,-\,A_{\rm pv},
\end{equation}
where the synonym $A_{\rm pv}$ denotes the spin asymmetry relating to the Coulombic field, i.e. $V(r)=V_T(r)$ in (\ref{2.1}). 
For low collision energies or small scattering angles (when $A_{\rm pv}$ keeps well away from zero), it is reasonable to define the
relative spin asymmetry change $dA_{\rm pv}$,
\begin{equation}\label{3.2}
dA_{\rm pv}\,=\,\frac{\Delta A_{\rm pv}}{A_{\rm pv}}.
\end{equation}

For dispersion, the relative spin asymmetry changes $dA_{\rm pv}^{\rm box}(L,\omega_L)$, pertaining to a given excited nuclear state with angular
momentum $L$ and excitation energy $\omega_L$, are additive,
\begin{equation}\label{3.3}
dA_{\rm pv}^{\rm box}\,\approx \, \sum_{L,\omega_L} dA_{\rm pv}^{\rm box}(L,\omega_L).
\end{equation}
This additivity is based on the fact that the relative cross section changes by dispersion are negligibly small (well below $10^{-6}$).

\begin{figure}
\vspace{-1.5cm}
\includegraphics[width=11cm]{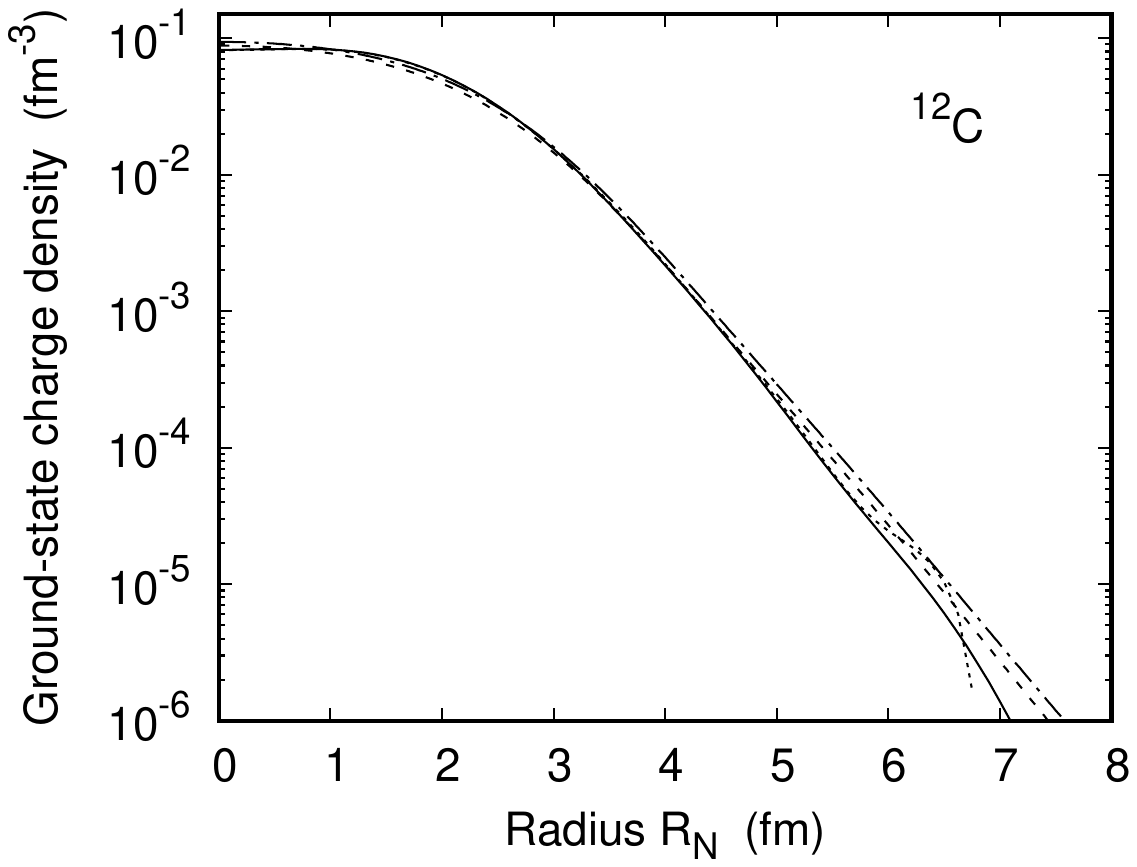}
\vspace{-1.5cm}
\caption{Ground-state charge density $\varrho_0$ of $^{12}$C as function of the distance $R_N$ from the nuclear center. Shown are the  parametrization Gauss (-------), the Fourier-Bessel fit $(\cdots\cdots$) and the result from the nuclear model $(-\cdot -\cdot -)$.
Included is the negative value of the weak density $\varrho_{\rm w}\;(----)$.
\label{fig2}}
\end{figure}

\begin{figure}
\vspace{-1.5cm}
\includegraphics[width=11cm]{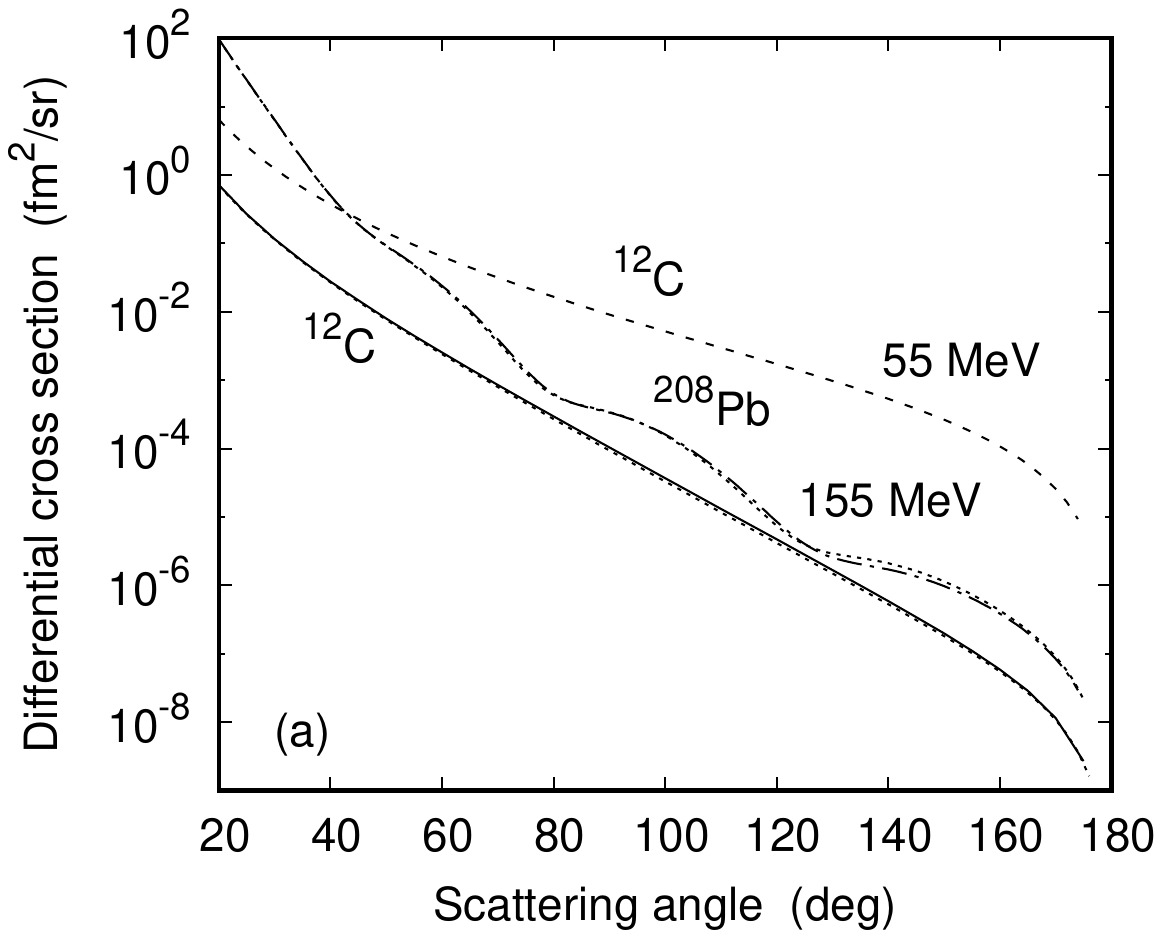}
\vspace{-1.5cm}
\vspace{-0.5cm}
\includegraphics[width=11cm]{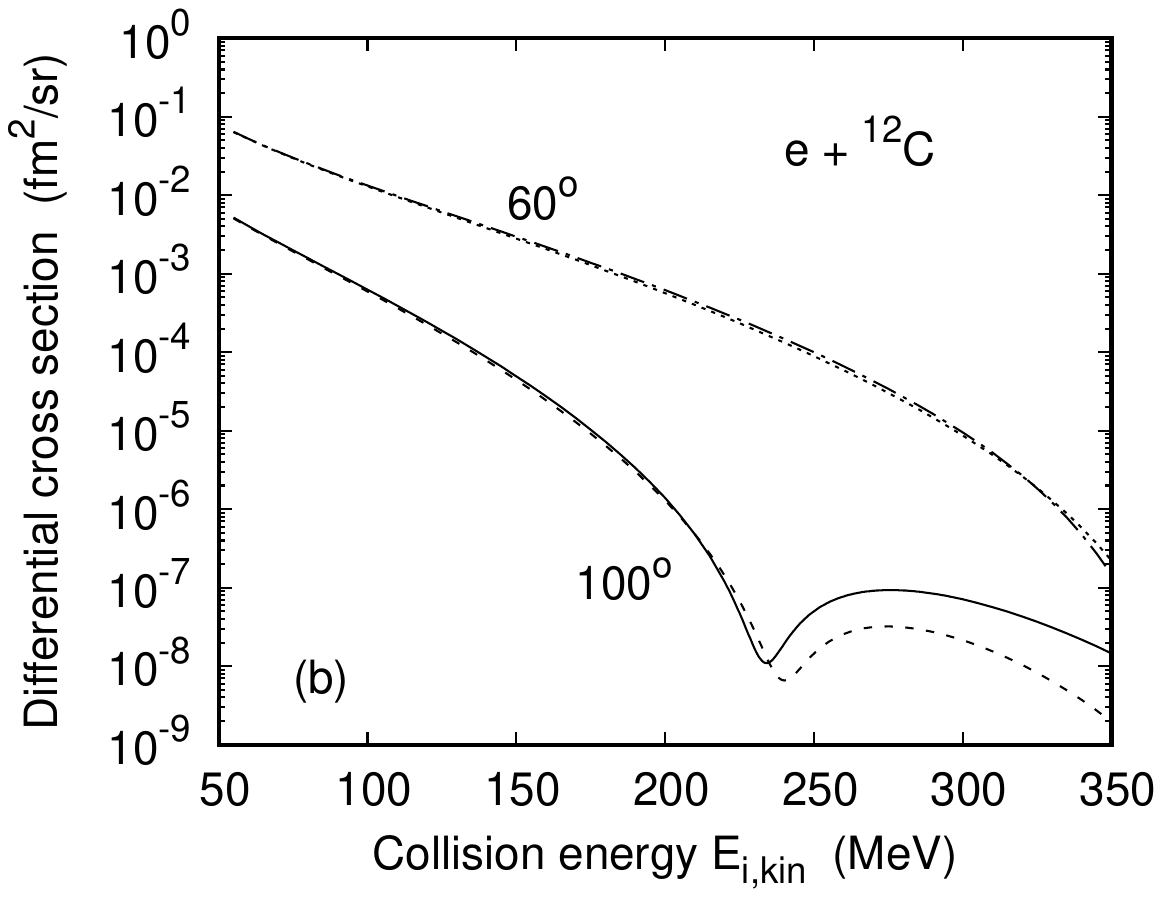}
\caption
{
Differential cross section $\frac{d\sigma_{\rm coul}}{d\Omega}$ for unpolarized electrons (a) scattering from $^{12}$C at 55 MeV ($----$) and at 155 MeV (----------), as well as from $^{208}$Pb at 155 MeV ($-\cdot -\cdot -)$ as function of the scattering angle $\theta$, using the charge distribution Gauss.
Included are the results from the numerical $\varrho_0$ for 155 MeV ($\cdots\cdots$).
(b) scattering from $^{12}$C at $\theta = 60^\circ \;(-\cdot -\cdot -)$ and $ 100^\circ$ (----------) as function of collision energy $E_{\rm i,kin}$, using the Gauss $\varrho_0$.
Included are the results from the numerical $\varrho_0$ for $60^\circ \; (\cdots\cdots)$ and $100^\circ \; (----)$.
\label{fig3}}
\end{figure}

\begin{figure}
\vspace{-1.5cm}
\includegraphics[width=11cm]{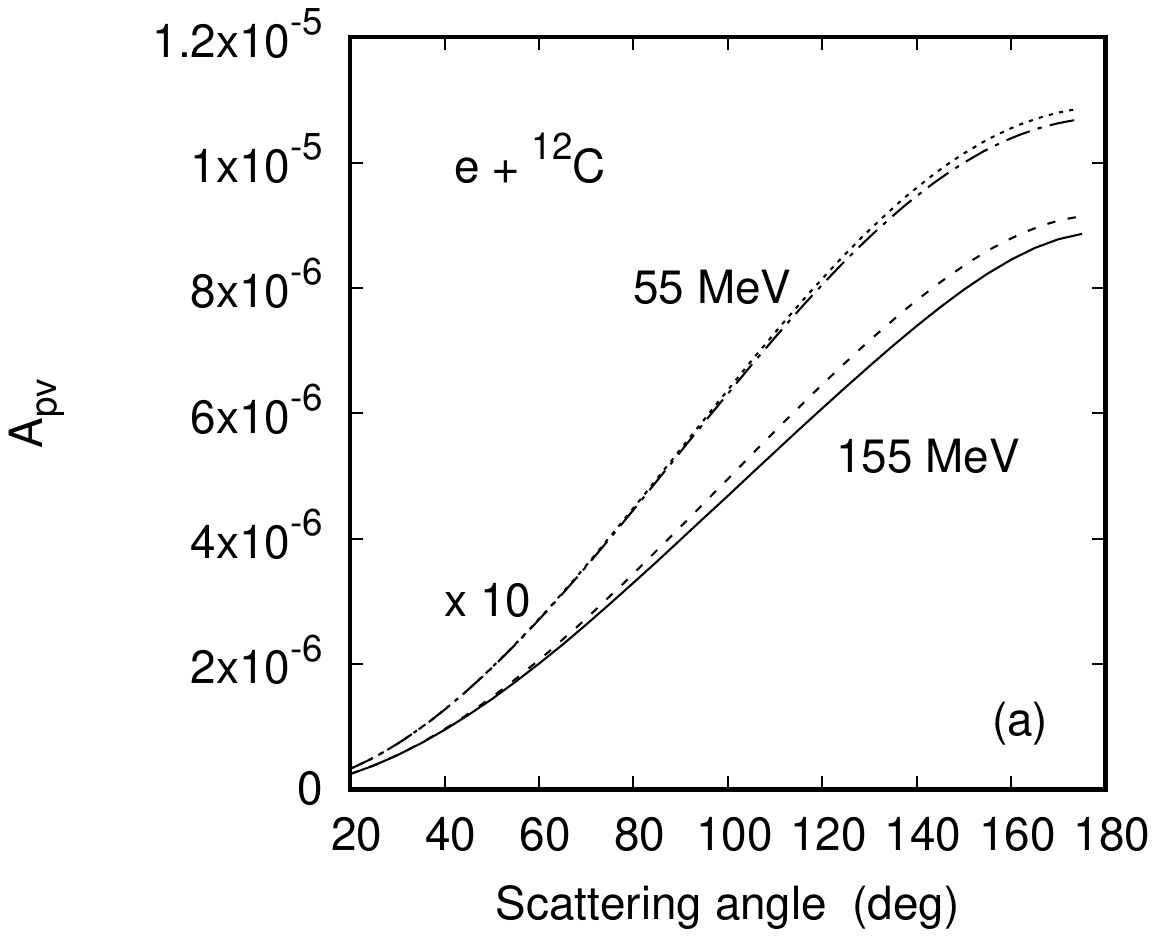}
\vspace{-1.5cm}
\vspace{-0.5cm}
\includegraphics[width=11cm]{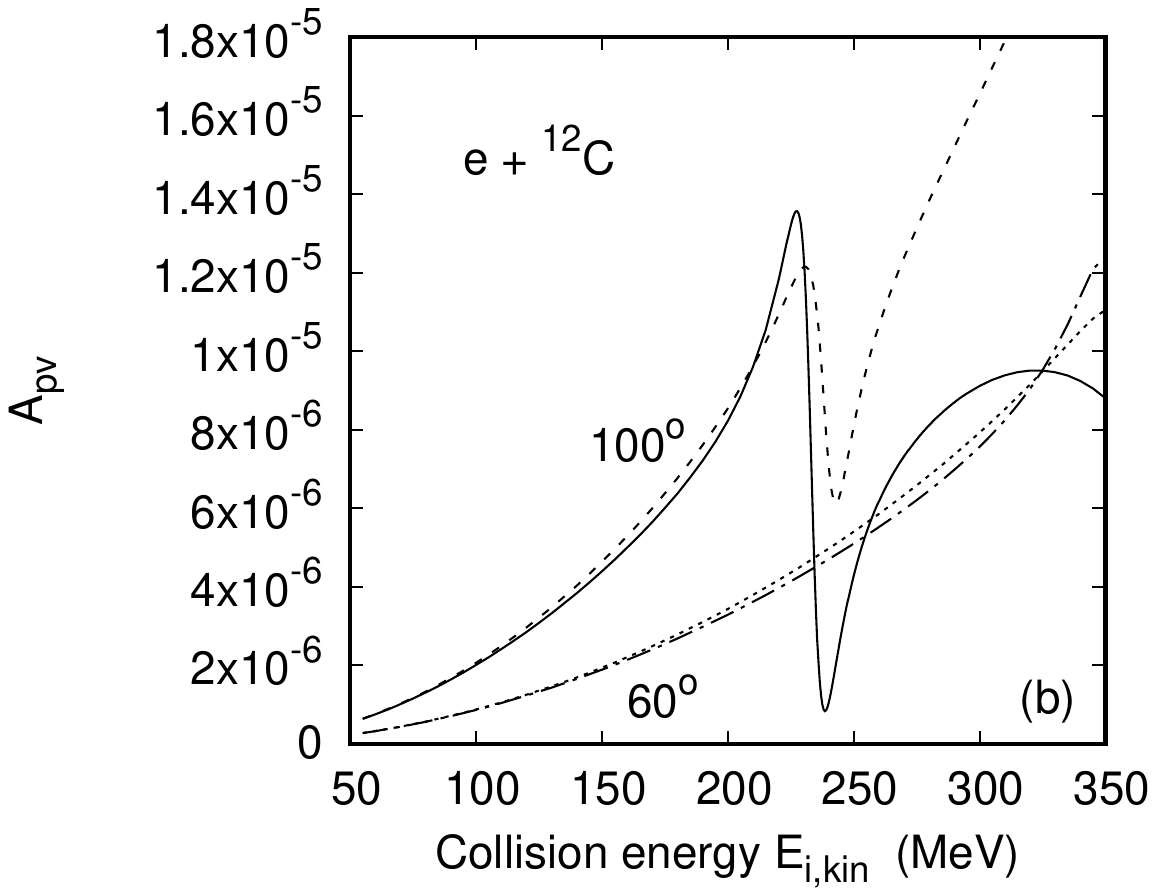}
\caption
{
Parity-violating asymmetry $A_{\rm pv}$ for electron scattering from $^{12}$C (a) at 55 MeV ($-\cdot -\cdot -$, multiplied by a factor of 10) and 155 MeV (----------) as function of scattering angle $\theta$
and (b) at $60^\circ \; (-\cdot -\cdot -)$ and $100^\circ$ (-------------) as function of collision energy, using the  charge distribution Gauss.
The results from the numerical charge density are also shown, at 55 MeV $(\cdots\cdots$, multiplied by 10) and 155 MeV $(----)$ in (a) and at $60^\circ \; (\cdots\cdots)$ and $100^\circ \;(-----)$ in (b).
\label{fig4}}
\end{figure}

\begin{figure}
\vspace{-1.5cm}
\includegraphics[width=11cm]{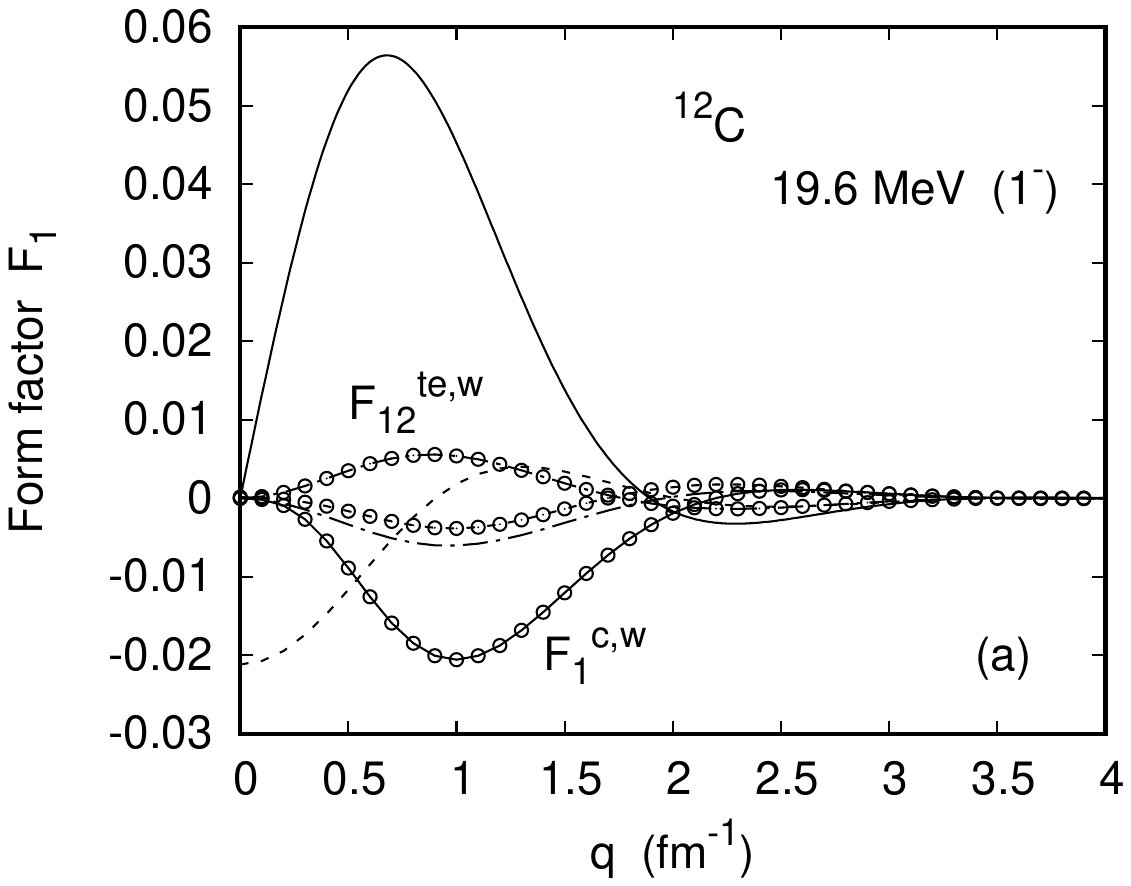}
\vspace{-1.5cm}
\vspace{-0.5cm}
\includegraphics[width=11cm]{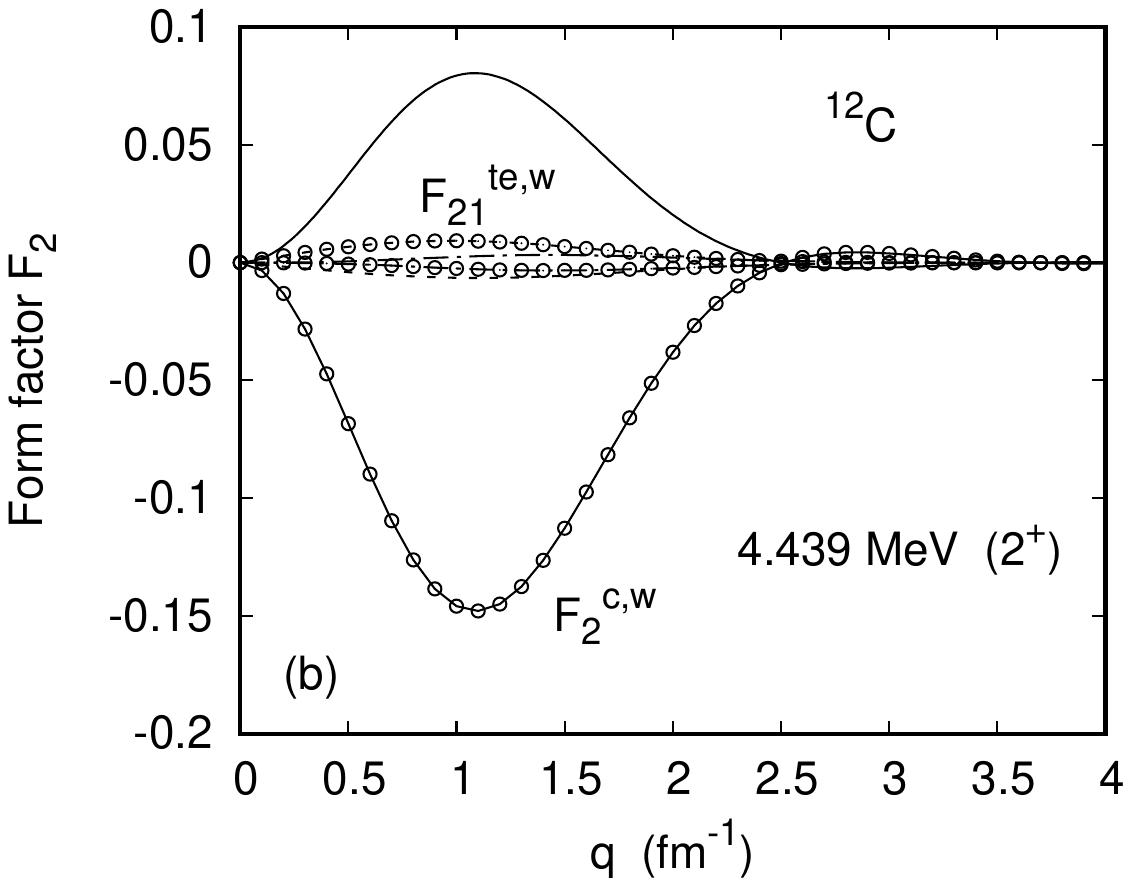}
\caption
{
$^{12}$C transition form factors as function of momentum transfer (a) for the isovector $1^-$ state at 19.6 MeV and (b) for the isoscalar $2^+$ state at 4.439 MeV.
Shown are the proton form factor $F_L^{\rm c}$ (-----------), the weak form factor $F_L^{\rm c,w}$ (--$\circ$--),
the transverse electric form factors $F_{L,L+1}^{\rm te}\; (-\cdot - \cdot -)$ and $F_{L,L-1}^{\rm te}\;(----)$
as well as their weak counterparts $F_{L,L+1}^{\rm te,w}\; (-\cdot -\circ - \cdot -)$ and $F_{L,L-1}^{\rm te,w}\; (--\circ --)$ for (a) $L=1$ and (b) $L=2$. 
\label{fig5}}
\end{figure}
When the two kinds of radiative corrections are of comparable size,
they have to be combined. However, 
 a simple addition is not possible since the relative cross section change $\Delta \sigma^{\rm QED} =\frac{d\sigma^{\rm QED}}{d\Omega} / \frac{d\sigma_{\rm coul}}{d\Omega} -1$ by the vacuum polarization and the vs effect is in general large (up to -20\%). Therefore the weak spin asymmetry which  includes the radiative corrections has to be calculated from
\begin{equation}\label{3.4}
A_{\rm pv}^{\rm box+QED} \,\approx\, A_{\rm pv}^{\rm QED} + A_{\rm pv} \cdot dA_{\rm pv}^{\rm box} \,\frac{1}{1+\Delta \sigma^{\rm QED}},
\end{equation}
and correspondingly its relative change,
\begin{equation}\label{3.5}
dA_{\rm pv}^{\rm box+QED}\,\approx\, dA_{\rm pv}^{\rm QED}\,+\,dA_{\rm pv}^{\rm box} \cdot \frac{1}{1+\Delta \sigma^{\rm QED}}.
\end{equation}
Two spin-zero nuclei, $^{12}$C and $^{208}$Pb, are considered in what follows.

\subsection{The $^{12}$C nucleus}
For the ground-state charge distribution $\varrho_0$, from which the nuclear potential $V_T$ is generated, three different prescriptions were considered:
a  parametrization (Gauss) with a sum of Gaussians \cite{deV}, a Fourier-Bessel (FB) parametrization \cite{deV}, both obtained from a fit to elastic scattering data, and a numerical $\varrho_0$ calculated from the SkM* nuclear model \cite{skm}. They are displayed in Fig.~\ref{fig2}.
The shapes of the FB and the Gauss parametrization are nearly identical (up to 5.5 fm), with a minimum near $R_N=0$.
However, there are considerable deviations between these two densities and the theoretical one.

Fig.~\ref{fig3} shows the angular  and energy distribution of the differential cross section for the elastic scattering of unpolarized electrons, as obtained by means of the phase-shift analysis \cite{Lan}.
At low collision energies $E_{\rm i,kin} =E_i-c^2$ (Fig.~\ref{fig3}a) the difference arising from the choice of the Gauss $\varrho_0$ and the theoretical $\varrho_0$ is small,
mostly below 10\% at 155 MeV and less ($\lesssim$ 3\%) at 55 MeV. However, in the vicinity of the first diffraction minimum (at a momentum transfer $q\approx 1.84 $ fm$^{-1}$, Fig.~\ref{fig3}b)
these deviations are formidable, the more so, the higher the collision energy.
The above momentum transfer corresponds to a mean electron-nucleus distance $R_N\approx q^{-1}=0.54$ fm where the respective choices of $\varrho_0$ are very different.

\begin{figure}
\vspace{-1.5cm}
\includegraphics[width=11cm]{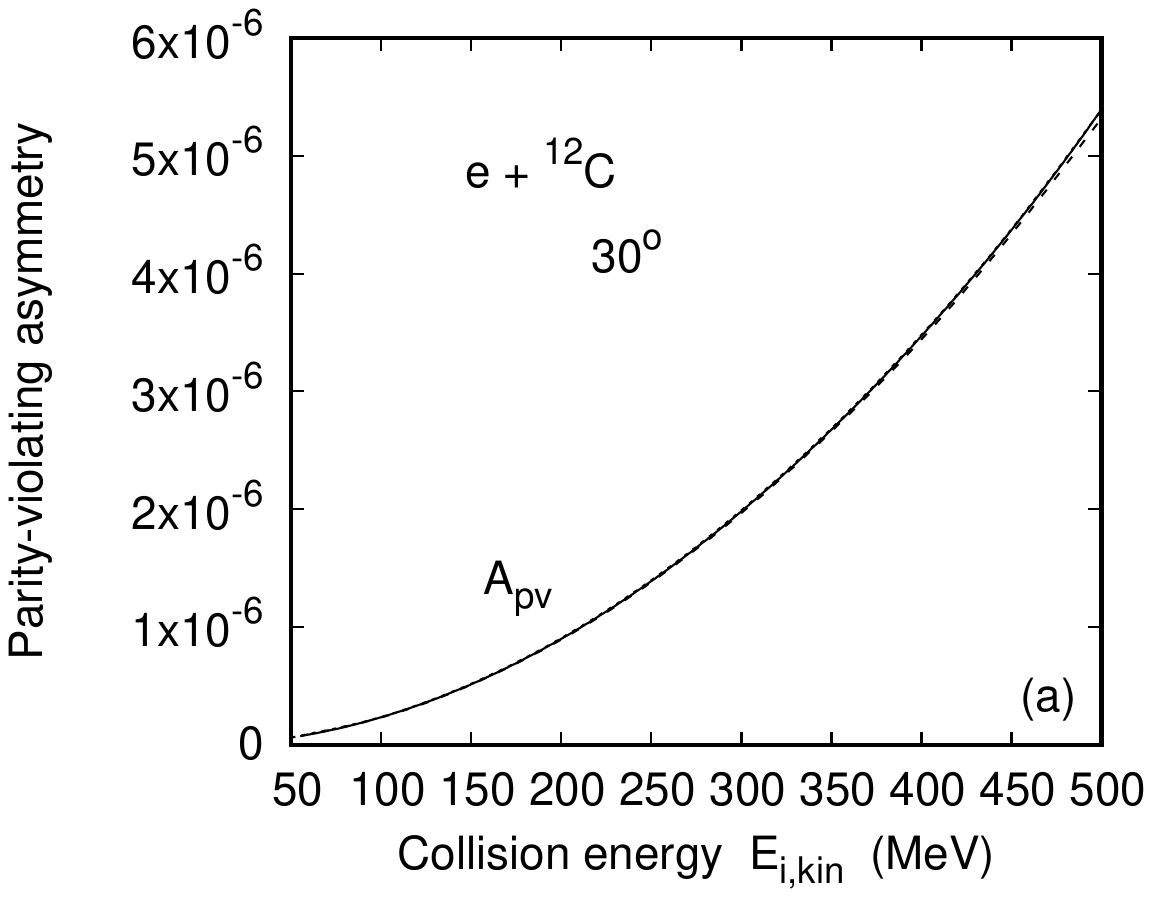}
\vspace{-1.5cm}
\vspace{-0.5cm}
\includegraphics[width=11cm]{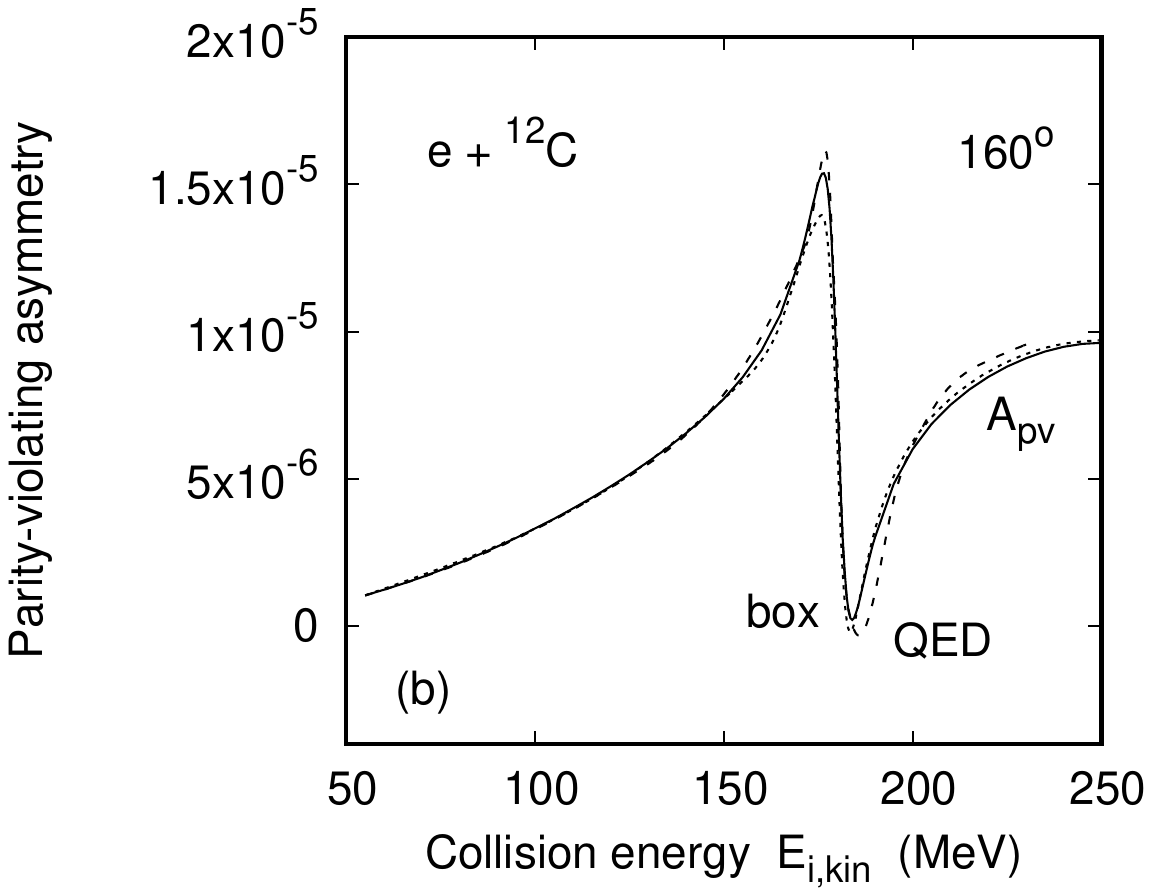}
\caption
{
Parity-violating asymmetry $A_{\rm pv} $ (----------) in $e+^{12}$C collisions, corrected for QED effects $(A_{\rm pv}^{\rm QED},\;----)$ and for dispersion $(A_{\rm pv}^{\rm box},\;\cdots\cdots)$ (a) at $30^\circ$ and (b) at $160^\circ$ as function of collision energy. In (a), $A_{\rm pv}^{\rm box}$ coincides with $A_{\rm pv}$.
\label{fig6}}
\end{figure}

The weak density $\varrho_{\rm w}$ is calculated from (\ref{3.6a})  as predicted by the SkM$^*$ \cite{skm}.
The weak charge  of a nucleus with $N$ neutrons and $Z$ protons  as defined by \cite{Ho98}
\begin{equation}\label{3.6}
Q_{\rm w}^{N,Z}=  \int d\bfr \,\varrho_{\rm w}(r) = N Q_n^{\rm w} +Z Q_p^{\rm w}
\end{equation}
is for $^{12}$C $-5.55$ at tree level and  $-5.51$ taking into account radiative corrections to $Q_p^{\rm w}=0.0719\pm0.0045$ and $Q_n^{\rm w}=-0.9890\pm0.0007$ \cite{Nav24}. 
The absolute value of $\varrho_{\rm w}$ for  ${}^{12}$C is depicted in Fig.~\ref{fig2}. Due to the fact that it is an $N=Z$ nucleus, the neutron and proton densities are very similar and thus, as can be seen, the electric and weak charge density distributions are also very similar in magnitude but have opposite signs. 
While our calculations are performed with $Q_{\rm w}^{N,Z}=-5.50$, a renormalization by a factor 5.55/5.50 reduces $A_{\rm pv}$ by up to 5\% at a collision energy of 155 MeV and angles up to $175^\circ$.

The influence of  $\varrho_0$ on the spin asymmetry $A_{\rm pv}$ is displayed in Fig.~\ref{fig4}.
The deviations are larger than for the differential cross section, particularly in the region of the diffraction minimum where $A_{\rm pv}$ shows a large oscillatory structure (Fig.~\ref{fig4}b).
Beyond the diffraction minimum even the shape of the energy distribution  differs strongly.
This sensitivity to the model densities at higher momentum transfers has to be kept in mind as a much larger source of uncertainty than the radiative corrections.
In the subsequent calculations the  representation Gauss of the ground-state charge density is used.

\begin{figure}
\vspace{-1.5cm}
\includegraphics[width=11cm]{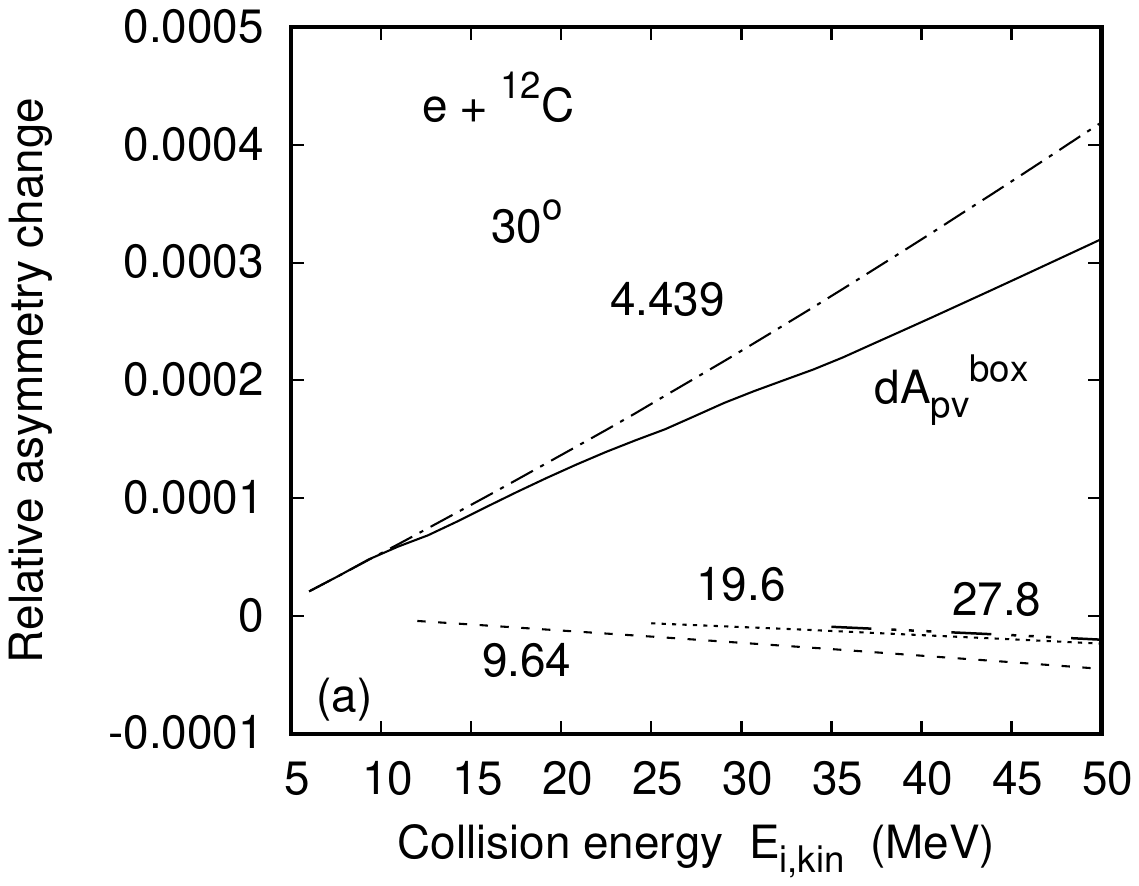}
\vspace{-1.5cm}
\vspace{-0.5cm}
\includegraphics[width=11cm]{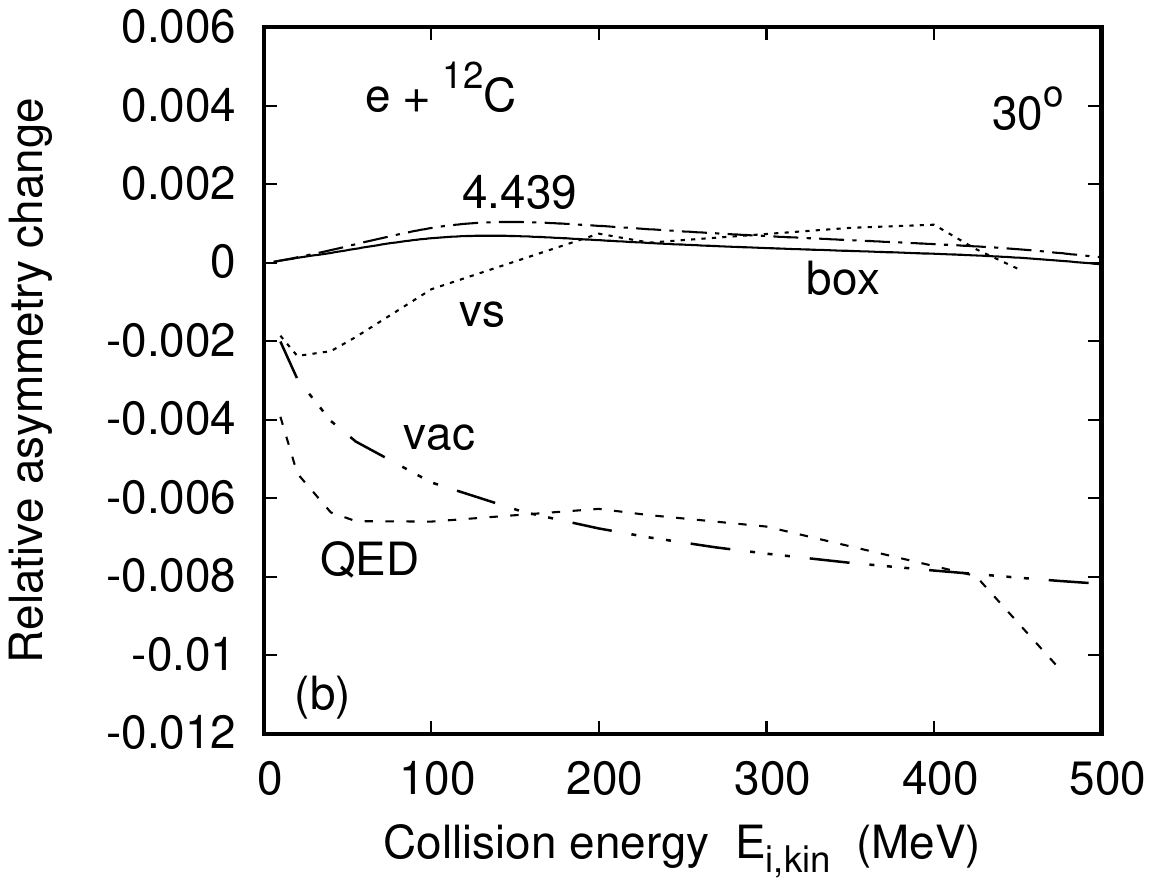}
\caption
{
Relative asymmetry change $dA_{\rm pv}$ in $e +^{12}$C collisions at $\theta = 30^\circ$ as function of collision energy. Shown is $dA_{\rm pv}^{\rm box}$ (summed over all excited states, ----------)
and $dA_{\rm pv}^{\rm box}(L,\omega_L)$ for the $2^+$ state at 4.439 MeV $(-\cdot -\cdot -)$.
Included in (a) is $dA_{\rm pv}^{\rm box}(L,\omega_L)$ for the $3^-$ state at 9.64 MeV $(----)$ and for the $1^-$ states at 19.6 MeV $(\cdots\cdots)$ and 27.8 MeV $(-\cdots -)$.
In (b), the change by the vertex plus self-energy correction $(\cdots\cdots)$, by the vacuum polarization $(-\cdots -)$ and by both QED effects ($dA_{\rm pv}^{\rm QED},\;----)$ is shown in addition. 
\label{fig7}}
\end{figure}

Concerning  the dispersive corrections,  we have considered the most relevant excited states with angular momentum $L\leq 3$ and excitation energy $\omega_L <30$ MeV as predicted by the SkM* effective interaction \cite{skm} solved via the Hartree-Fock plus Random Phase Approximation \cite{CM21}. The six selected excited states are: three $1^-$ states at 19.6 (isovector), 20.9 (isoscalar) and 27.8 MeV (isoscalar); two $2^+$ states
at 4.439 (isoscalar) and 9.84 MeV (isovector); and one $3^-$ state at 9.64 MeV (isoscalar).
The proton and neutron transition densities are calculated consistently within the same nuclear model.
The corresponding transition form factors together with their weak counterparts are displayed in Fig.~\ref{fig5} for the  19.6 MeV $(1^-$) and the  4.439 MeV ($2^+$) state.
The isoscalar states are characterized by a large weak charge form factor (Fig.~\ref{fig5}b), while for the isovector states the magnetic weak
form factors gain importance (Fig.~\ref{fig5}a).

The influence of the radiative effects on the weak spin asymmetry is shown in Fig.~\ref{fig6} at a forward and a backward scattering angle.
At small angles, dispersion is tiny and also the QED perturbation is mostly negligible (Fig.~\ref{fig6}a), while near and above the diffraction minimum, accessible at the larger angles, there is a sizeable influence of both QED and dispersion (Fig.~\ref{fig6}b).

Let us now consider the relative  changes by the QED effects $(d A_{\rm pv}^{\rm QED})$ and by dispersion $(d A_{\rm pv}^{\rm box}$). 
Their energy distribution at $30^\circ$ is displayed in Fig.~\ref{fig7}. It is seen that both radiative effects vanish when the collision energy  tends to zero.
In Fig.~\ref{fig7}a the separate dispersion results are shown for the 19.6 and 27.8 MeV dipole, the 4.439 MeV quadrupole and the 9.64 MeV octupole excitation (whereas the 20.9 
MeV and the 9.84 MeV contributions can be neglected).
It follows that the main contribution to dispersion is due to the lowest quadrupole excitation, while the dipole states are of minor importance.
This is opposite to the results found for the Sherman function where dispersion originates predominantly from the dipole states,
but that could also be due to a different choice of model \cite{Jaku22} for the dipole transition densities.
The present result can be understood in terms of the form factors (Fig.~\ref{fig5}).
While for the $2^+$ state the weak charge form factor is enhanced by a factor of two as compared to $F_L^c$,
it is reduced by a factor of three in the case of the $1^-$ state.
Also the magnetic weak form factors for this state are smaller than their protonic counterparts.
We note that the above-mentioned renormalization of $\varrho_{\rm w}$ (by the factor 5.55/5.50) changes $\Delta A_{\rm pv}^{\rm box}$ by at most 0.1\%.

\begin{figure}
\vspace{-1.5cm}
\includegraphics[width=11cm]{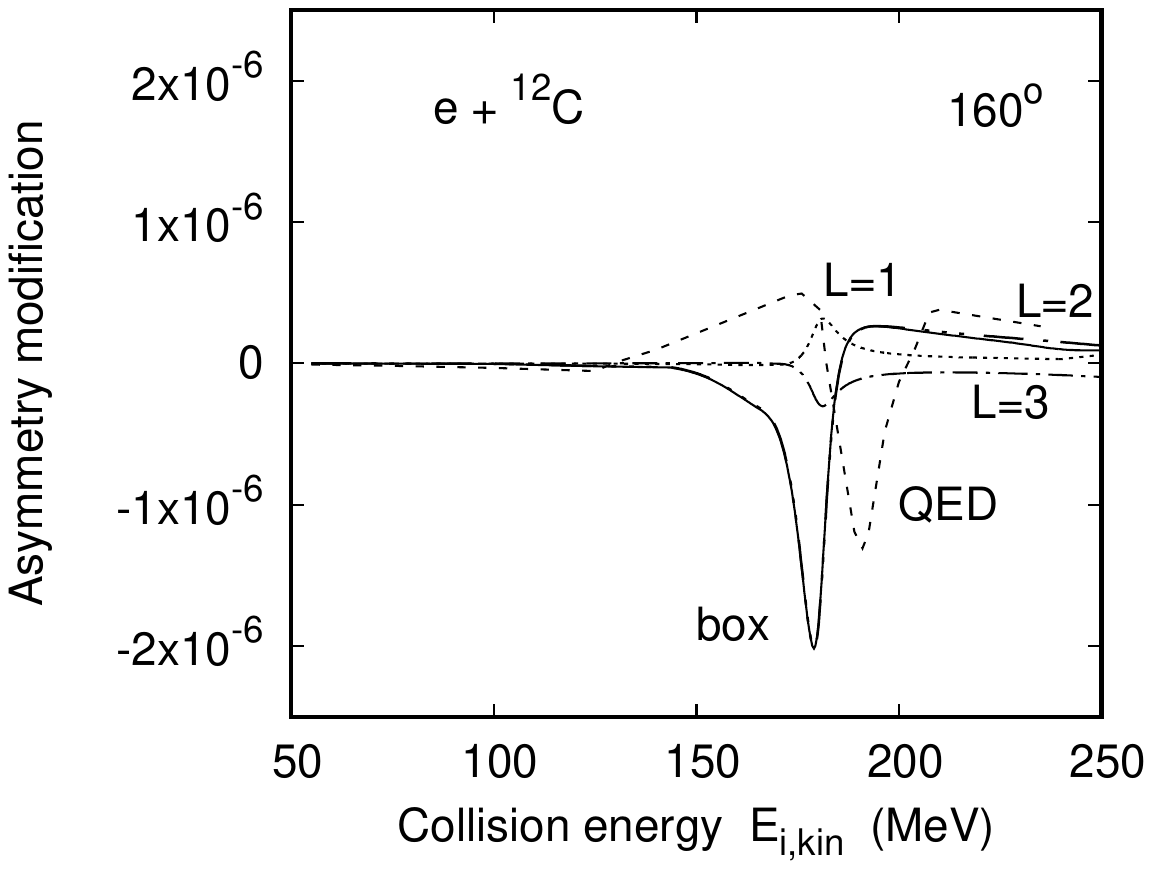}
\vspace{-1.5cm}
\caption
{
Spin asymmetry modification $\Delta A_{\rm pv}$ for electrons colliding with $^{12}$C at $160^\circ$ as function of collision energy. Shown are the dispersive modifications $\Delta A_{\rm pv}^{\rm box} = A_{\rm pv}^{\rm box} -A_{\rm pv}$ when summed over the dipole states $(\cdots\cdots)$, over the quadrupole states $(-\cdots - \cdots -)$, over the octupole states $(-\cdot -\cdot -)$ as well as over $L\leq 3$ (-------------).
Included is the asymmetry modification by the QED effects, $\Delta A_{\rm pv}^{\rm QED} = A_{\rm pv}^{\rm QED} -A_{\rm pv}\;(----)$.
\label{fig8}}
\end{figure}

The correction of $A_{\rm pv}$ by the QED effects (Fig.~\ref{fig7}b) is more important than dispersion at this forward angle.
The vector vs correction leads to a large effect on $A_{\rm pv}$, but is mostly compensated by the axial-vector vs contribution, with a net effect well below 1\%.
At  backward angles, however, dispersion and QED effects are similar in magnitude. This is shown for $160^\circ$ in Fig.~\ref{fig8}.
Due to the diffraction oscillations of $A_{\rm pv}$ at this angle, where the weak spin asymmetry nearly vanishes, the definition of the relative spin asymmetry change is no longer meaningful.
Therefore, the spin asymmetry modifications $\Delta A_{\rm pv}^{\rm box}$ and $\Delta A_{\rm pv}^{\rm QED}$ from (\ref{3.1}) are displayed instead.
The figure indicates
 that dispersion is nearly exclusively due to the lowest quadrupole state, showing a large excursion  near the location of the diffraction minimum.
The contribution from dipole and octupole states mostly cancel each other,
even for still higher energies.

\begin{figure}
\vspace{-1.5cm}
\includegraphics[width=11cm]{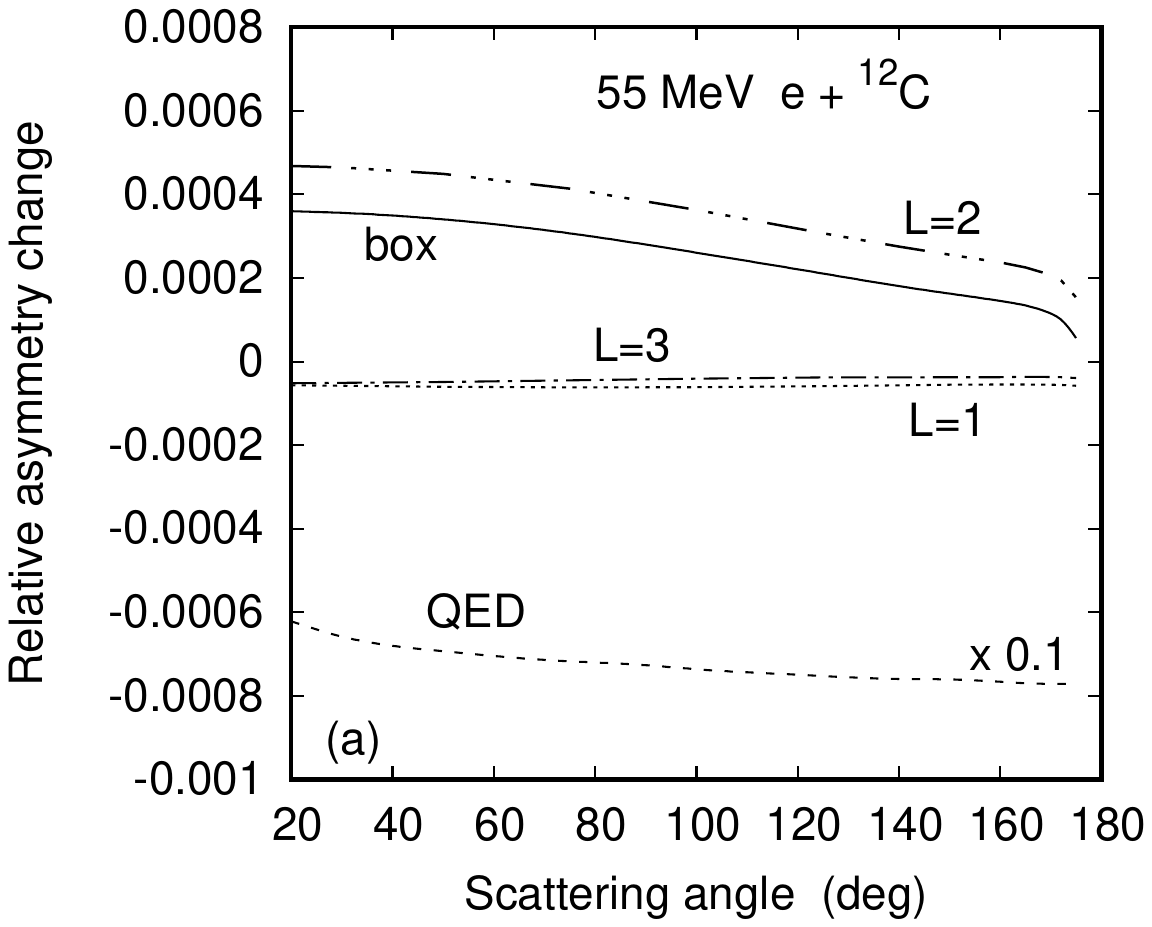}
\vspace{-1.5cm}
\vspace{-0.5cm}
\includegraphics[width=11cm]{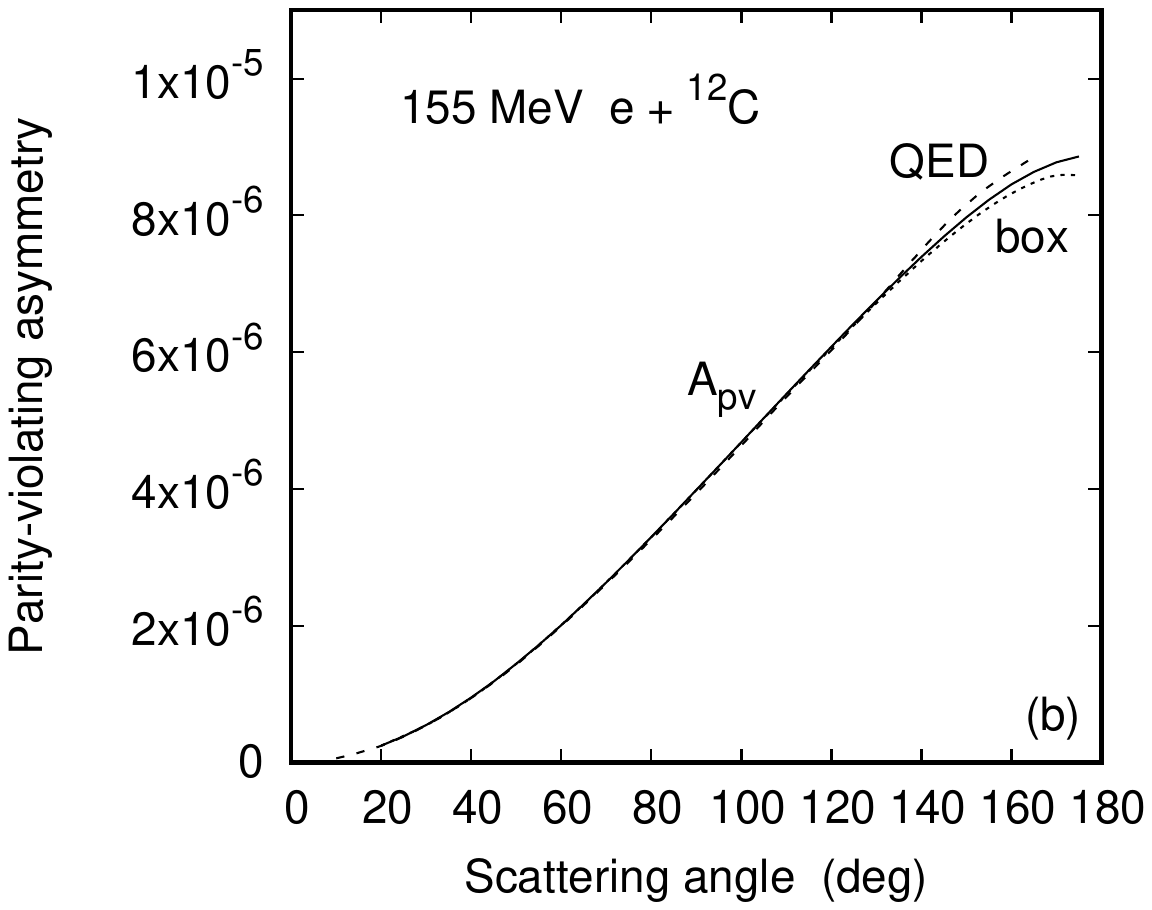}
\caption
{
Parity-violating spin asymmetry and its radiative corrections  in  $e+^{12}$C collisions (a) at 55 MeV and (b) at 155 MeV as function of scattering angle $\theta$.
Shown in (a) are the dispersive changes $dA_{\rm pv}^{\rm box} $ when summed over the dipole states $(\cdots\cdots)$, over the quadrupole states $(-\cdots - \cdots -)$, over the octupole states $(-\cdot -\cdot -)$ as well as over $L\leq 3$ (------------).
Included is the change $dA_{\rm pv}^{\rm QED}$ by the QED effects $(----$, reduced by a factor of 0.1).
Shown in (b) is $A_{\rm pv}$ (--------), $A_{\rm pv}^{\rm QED} (----)$ and $A_{\rm pv}^{\rm box}\;(\cdots\cdots)$. 
\label{fig9}}
\end{figure}

The angular distribution of the parity-violating asymmetry and its radiative changes for the two energies of 55 and 155 MeV is depicted in Fig.~\ref{fig9}.
While the QED effects are strongly  dominating at 55 MeV for all angles (Fig.~\ref{fig9}a),  $dA_{\rm pv}^{\rm box}$ gains importance at the higher energy
 for backward angles.

The $L=2$ contribution is for 155 MeV dominating at all angles, while the changes by dipole and octupole excitations are at most in the permille region.

We have also studied the radiative corrections in the high-energy regime.  As a test case we have considered an energy of  953 MeV and a scattering angle of $4.7^\circ$, a
 geometry which has been selected for $A_{\rm pv}$ precision experiments on the lead nucleus \cite{Ad21}.
For $A_{\rm pv}$, being equal to $5.033 \times 10^{-7}$, the relative change by the QED effects is $
d A_{\rm pv}^{\rm QED} \approx -0.51\%$ and by dispersion $d A_{\rm pv}^{\rm box} =-0.02\%.$
Thus the dispersive effects from low-lying excited states can safely be neglected at such geometries and incident electron energies, and also the QED corrections play only a minor role.

\subsection{The $^{208}$Pb nucleus}

\begin{figure}
\vspace{-1.5cm}
\includegraphics[width=11cm]{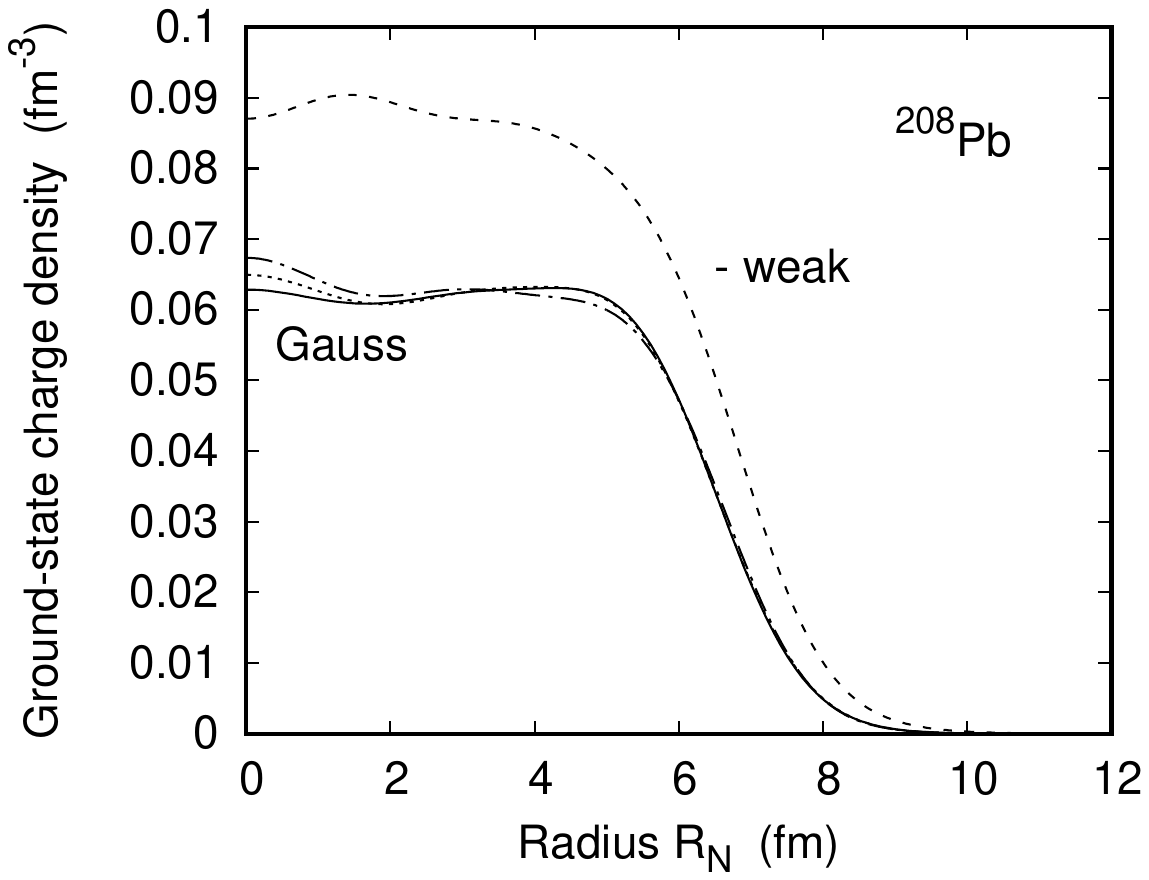}
\vspace{-1.5cm}
\caption
{
Ground-state charge density $\varrho_0$ of $^{208}$Pb as function of the distance from the nuclear center.
Shown are the parametrization Gauss (----------), the FB fit $(\cdots\cdots$) and the result from the nuclear model $(-\cdot -\cdot -$).
Included is the negative of the weak density $\varrho_{\rm w}\;(----)$. \label{fig10}}
\end{figure}

\begin{figure}
\vspace{-1.5cm}
\includegraphics[width=11cm]{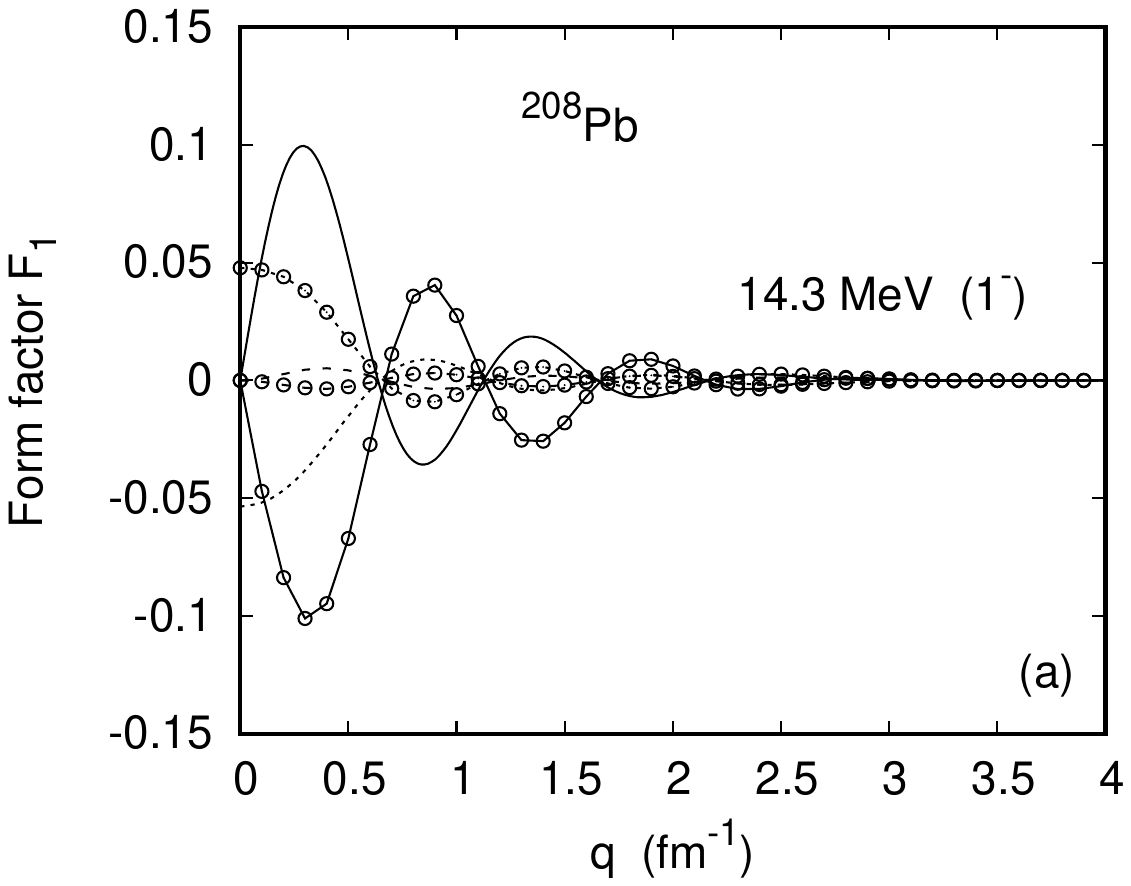}
\vspace{-1.5cm}
\vspace{-0.5cm}
\includegraphics[width=11cm]{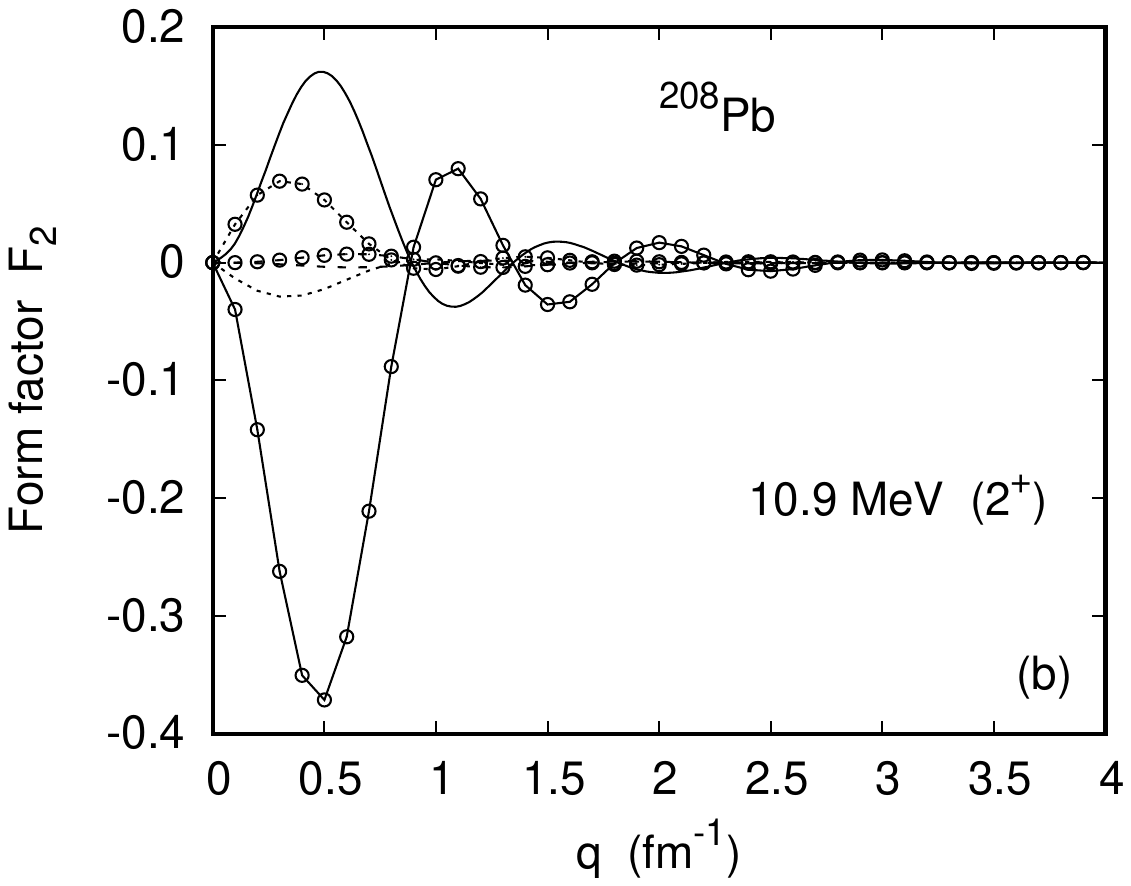}
\vspace{-1.5cm}
\includegraphics[width=11cm]{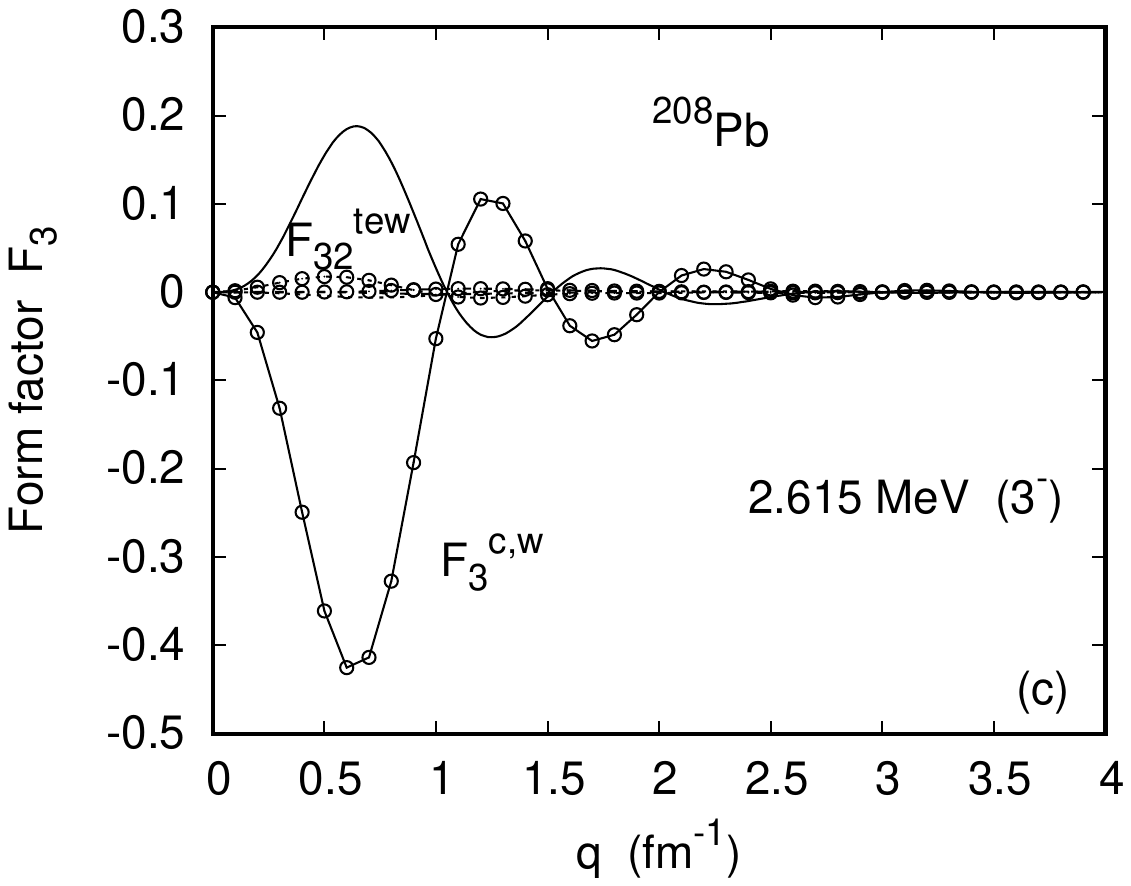}
\vspace{1.0cm}
\caption
{
Transition form factors as function of momentum transfer, (a) for the $^{208}$Pb isovector $1^-$ state at 14.3 MeV, (b) for the isoscalar $2^+$ state at 10.9 MeV and (c) for the isoscalar $3^-$ state at 2.615 MeV.
Shown are the proton form factor $F_L^{\rm c}$ (------------), the weak form factor $F_L^{\rm c,w}$ (---$\circ$---),
the transverse electric form factors $F_{L,L+1}^{\rm te}\;(----)$ and $F_{L,L-1}^{\rm te}\;(\cdots\cdots)$ as well as their weak counterparts $F_{L,L+1}^{\rm te,w}\;(--\circ --)$ and $F_{L.L-1}^{\rm te,w}\;(\cdots \circ \cdots)$ for (a) $L=1$, (b) $L=2$ and (c) $L=3$.
\label{fig11}}
\end{figure}

Like for the carbon nucleus, the  parametrization Gauss \cite{deV} of the experimentally determined nuclear ground-state charge distribution $\varrho_0$  is employed in the subsequent calculations.
A comparison with the Fourier-Bessel charge distribution \cite{deV} shows a basic agreement except for very small nuclear distances as seen in Fig.~\ref{fig10}.
In contrast, the numerically obtained $\varrho_0$ from the nuclear model SkP \cite{skp} deviates from the Gauss $\varrho_0$ in large regions of space.
The effect of these deviations on the differential cross section for elastic scattering is negligibly small, as displayed in Fig.~\ref{fig3}a for a collision energy of 155 MeV, except at the larger angles.
The weak density for $^{208}$Pb, normalized to $Q_{\rm w}=-118.76$ (which includes the universal radiative corrections \cite{Nav24}), is also shown in Fig.~\ref{fig10},
again with a global minus sign for the sake of shape comparison.
In contrast to the results for $^{12}$C, the absolute value of $\varrho_{\rm w}$ is about 30\% higher than $\varrho_0$ for $R_N\lesssim 5$ fm, and even more in the outer region due to the large neutron skin of the lead nucleus.

In order to account for dispersion, 10 dominant excited states with $L\leq 3$ and $\omega_L <30$ MeV are considered.
These are the five $1^-$ states at energies 5.512 MeV (isoscalar), and at 12.6, 14.3, 14.6 and 15.3 MeV (isovector).
In addition, the three quadrupole states at 4.085 and 10.9 MeV (isoscalar) and at  21.6 MeV (isovector),
as well as two octupole states at 2.615 MeV (isoscalar) and 27.9 MeV (isovector) are included.
They are consistently calculated from the same (SkP) model \cite{skp}.
It turns out that the most  important ones are the 10.9 MeV $2^+$ and the 2.615 MeV $3^-$ states.
Their transition form factors are given in Fig.~\ref{fig11}, together with the corresponding weak form factors. It is seen that for both states the weak charge form factors ($F_L^{\rm c,w}$) are about a factor of two larger than the corresponding proton form factors ($F_L^{\rm c}$).
Like for the isoscalar states of $^{12}$C, the magnetic form factors are much smaller, particularly for the 2.615 MeV state.
For the sake of comparison, the form factors of the most relevant of the dipole states, the one at 14.3 MeV, are also shown (Fig.~\ref{fig11}a).
For all isovector $1^-$ states, proton and weak form factors have basically the same shape (but opposite sign).

\begin{figure}
\vspace{-1.5cm}
\includegraphics[width=11cm]{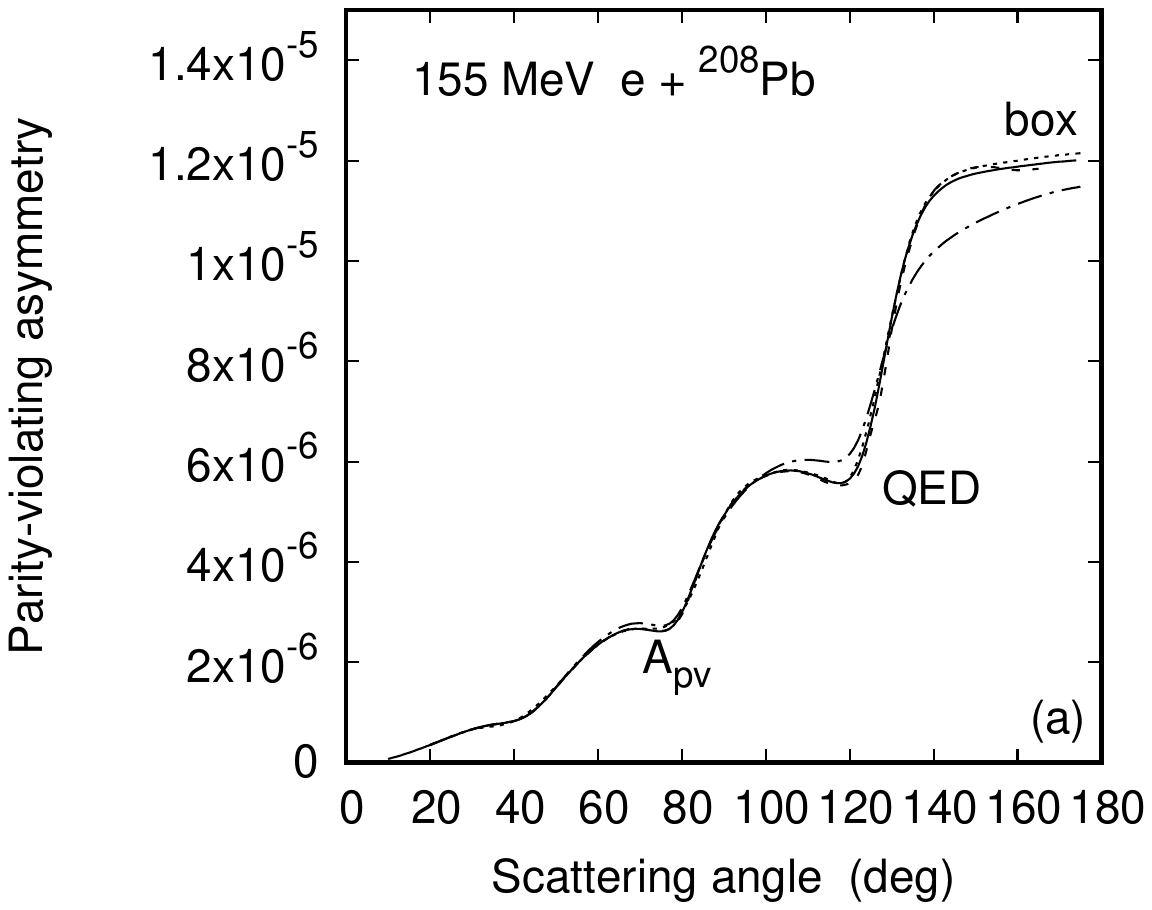}
\vspace{-1.5cm}
\vspace{-0.5cm}
\includegraphics[width=11cm]{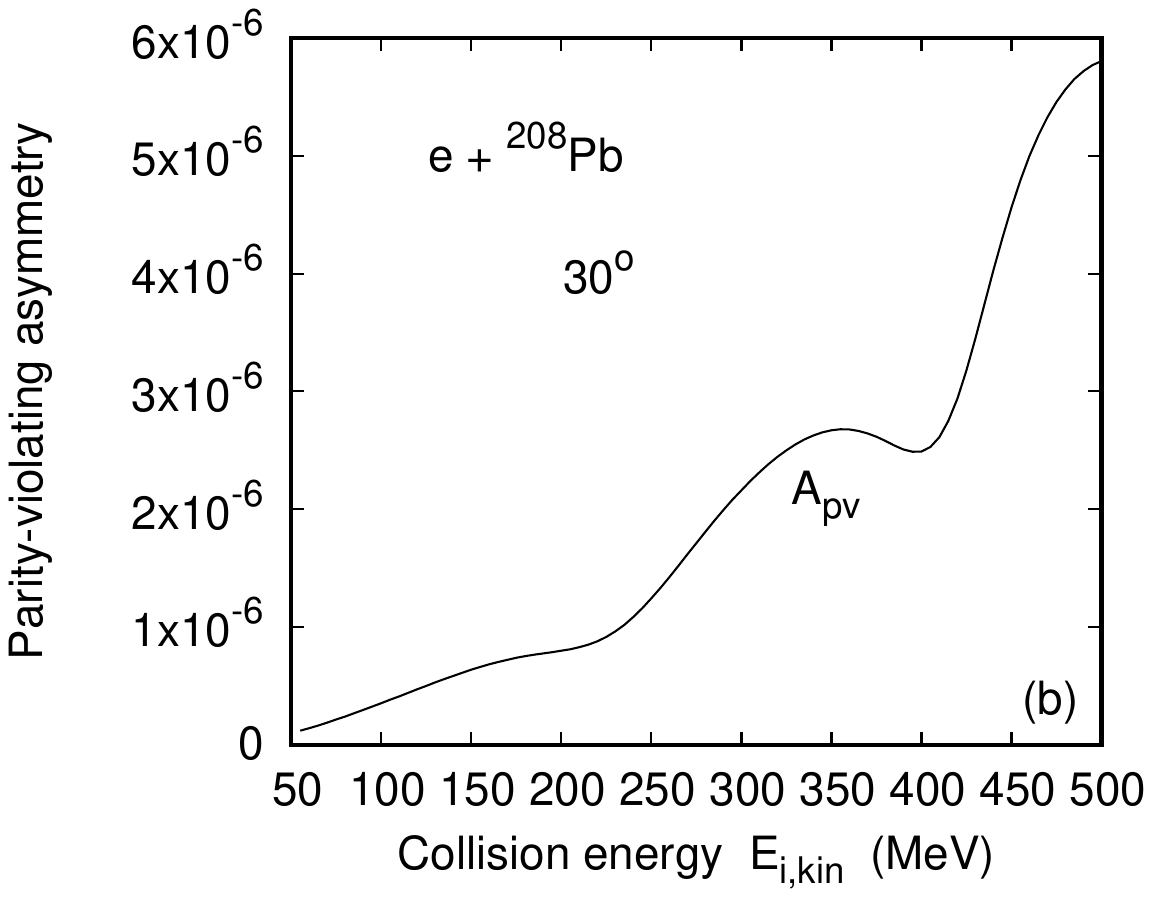}
\caption
{
Parity-violating asymmetry $A_{\rm pv}$ (-----------) in $e+^{208}$Pb collisions,  corrected for QED effects $(A_{\rm pv}^{\rm QED},\;----)$ and for dispersion $(A_{\rm pv}^{\rm box},\;\cdots\cdots)$
(a) at 155 MeV as function of scattering angle $\theta$ and (b) at $30^\circ$ as function of collision energy $E_{\rm i,kin}$.
In (a), also $A_{\rm pv}$ resulting from the numerical $\varrho_0$ is shown $(-\cdot - \cdot -)$.
In (b), $A_{\rm pv}^{\rm QED}$ and $A_{\rm pv}^{\rm box}$ coincide with $A_{\rm pv}$.
\label{fig12}}
\end{figure}

The parity-violating spin asymmetry together with its radiative corrections is shown in Fig.~\ref{fig12}. 
The angular distribution of $A_{\rm pv}$ at 155 MeV (Fig.~\ref{fig12}a) exhibits an oscillatory behaviour in concord with the diffraction structures of the differential cross section (see Fig.~\ref{fig3}a) which, however, 
are quite shallow.
The QED effects slightly reduce  the spin asymmetry, except around $140^\circ$.
Dispersion increases the spin asymmetry, well below 0.1\% for $\theta \lesssim 60^\circ$, but up to 1\% for the backward angles.

The behaviour of $A_{\rm pv}$ and its radiative corrections with energy is displayed in Fig.~\ref{fig12}b for the forward scattering angle of $30^\circ$.
The energy dependence of the diffraction structures is much alike their angular dependence, confirming that $A_{\rm pv}$ is basically dependent on momentum transfer which
increases in a similar way with angle and with energy.

Again the dispersive corrections enhance the weak spin asymmetry, but they are unimportant at this angle.
Also the QED effects are small throughout.

\begin{figure}
\vspace{-1.5cm}
\includegraphics[width=11cm]{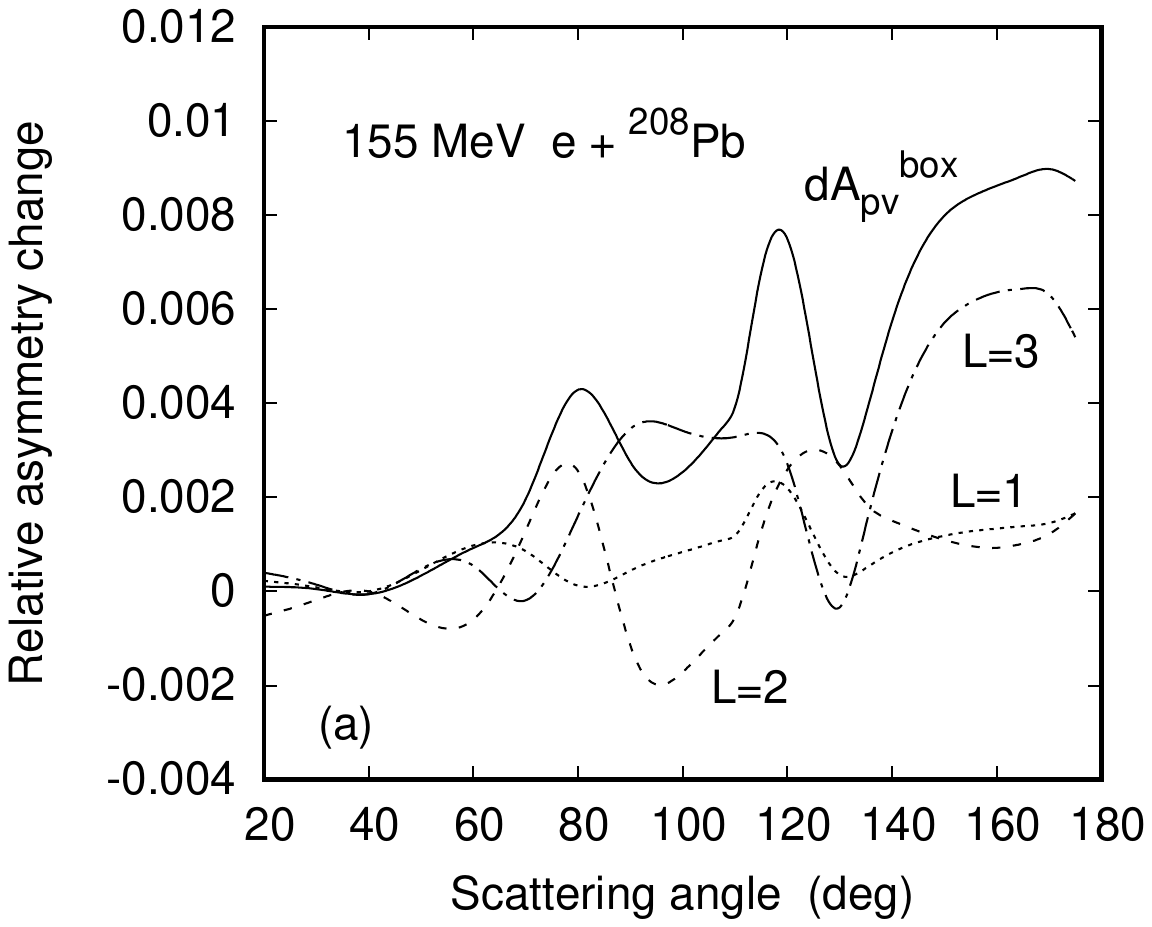}
\vspace{-1.5cm}
\vspace{-0.5cm}
\includegraphics[width=11cm]{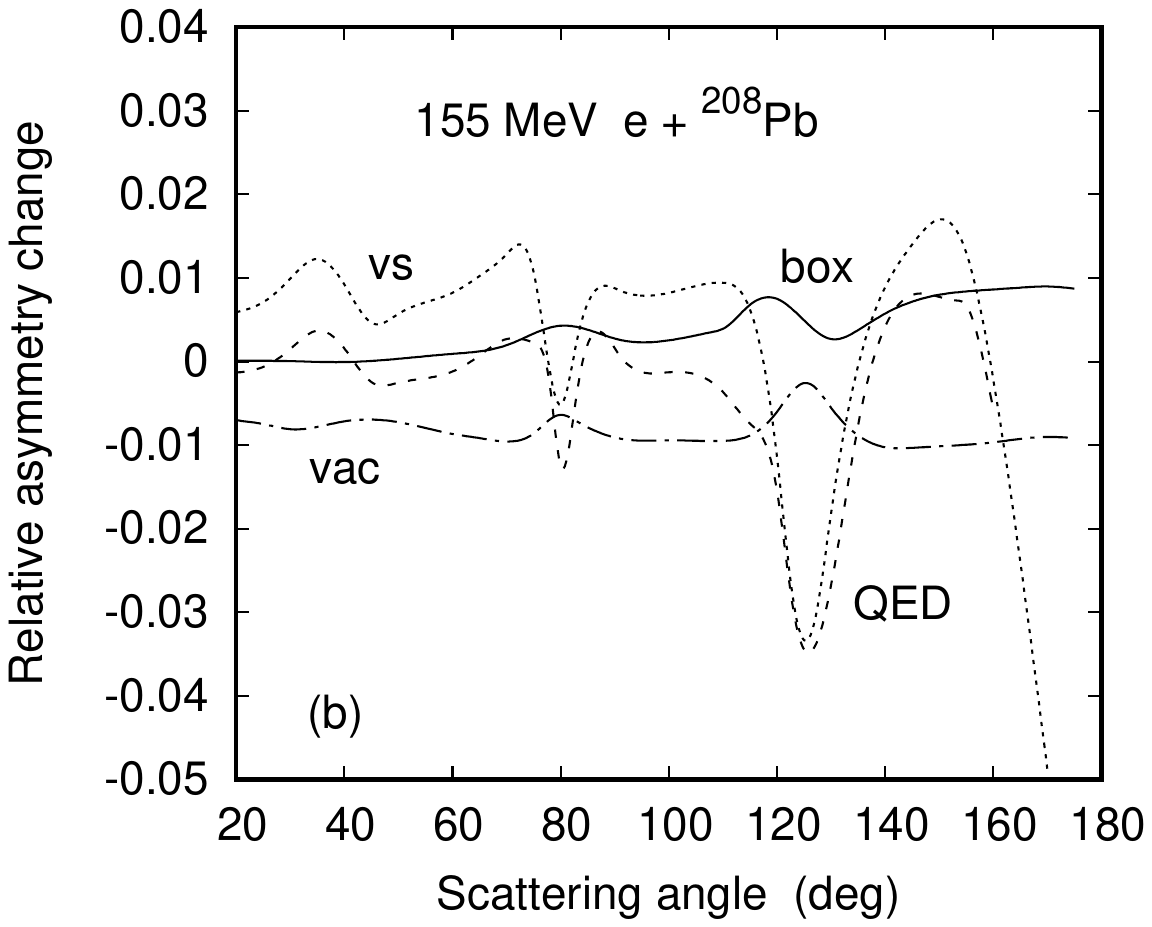}
\caption
{
Relative spin asymmetry change $dA_{\rm pv}$ in 155 MeV $e+^{208}$Pb collisions as function of scattering angle $\theta$.
Shown in (a) are the dispersive changes $dA_{\rm pv}^{\rm box}$ when summed over the dipole states $(\cdots\cdots)$,
over the quadrupole states $(----)$, over the octupole states $(-\cdot -\cdot -)$
and over all states with $L\leq 3$ ($dA_{\rm pv}^{\rm box}$, -------).
Shown in (b) are the QED changes by vacuum polarization $(-\cdot -\cdot -)$, by the vertex plus self-energy correction $(\cdots\cdots)$ and by both QED effects
$(dA_{\rm pv}^{\rm QED},\;----)$. Included is $dA_{\rm pv}^{\rm box}$ (-----------)  from (a).
\label{fig13}}
\end{figure}

Fig.~\ref{fig13} shows the relative change of $A_{\rm pv}$ by the radiative corrections at 155 MeV.
For dispersion (Fig.~\ref{fig13}a), the dominant contribution at the larger angles originates from the $L=2$ and $L=3$ excited states, with the 2.615 MeV state prevailing beyond $140^\circ$.
The dipole states determine the angular distribution only at angles below $60^\circ$, and this basically because the quadrupole and octupole contributions cancel each other in this angular region.

Dispersion increases globally with angle (Fig.13b). The structures seen in the angular distribution are conform with the oscillations of $A_{\rm pv}$ in Fig.~\ref{fig12}a.
Each minimum in $\frac{d\sigma_{\rm coul}}{d\Omega}$ or in $A_{\rm pv}$ produces a large excursion of both $dA_{\rm pv}^{\rm QED}$ and $dA_{\rm pv}^{\rm box}$.
Included in Fig.~\ref{fig13}b are the two constituents of $dA_{\rm pv}^{\rm QED}$, the change by vacuum polarization 
(amounting to about 1\%) and by the vs correction which is of opposite sign. The  QED corrections are  small in the forward hemisphere because vacuum polarization mostly compensates the vs correction. This latter correction is intrinsically small because also here the vector vs and the axial-vector vs contributions tend to cancel each other
except at the larger angles.

\begin{figure}
\vspace{-1.5cm}
\includegraphics[width=11cm]{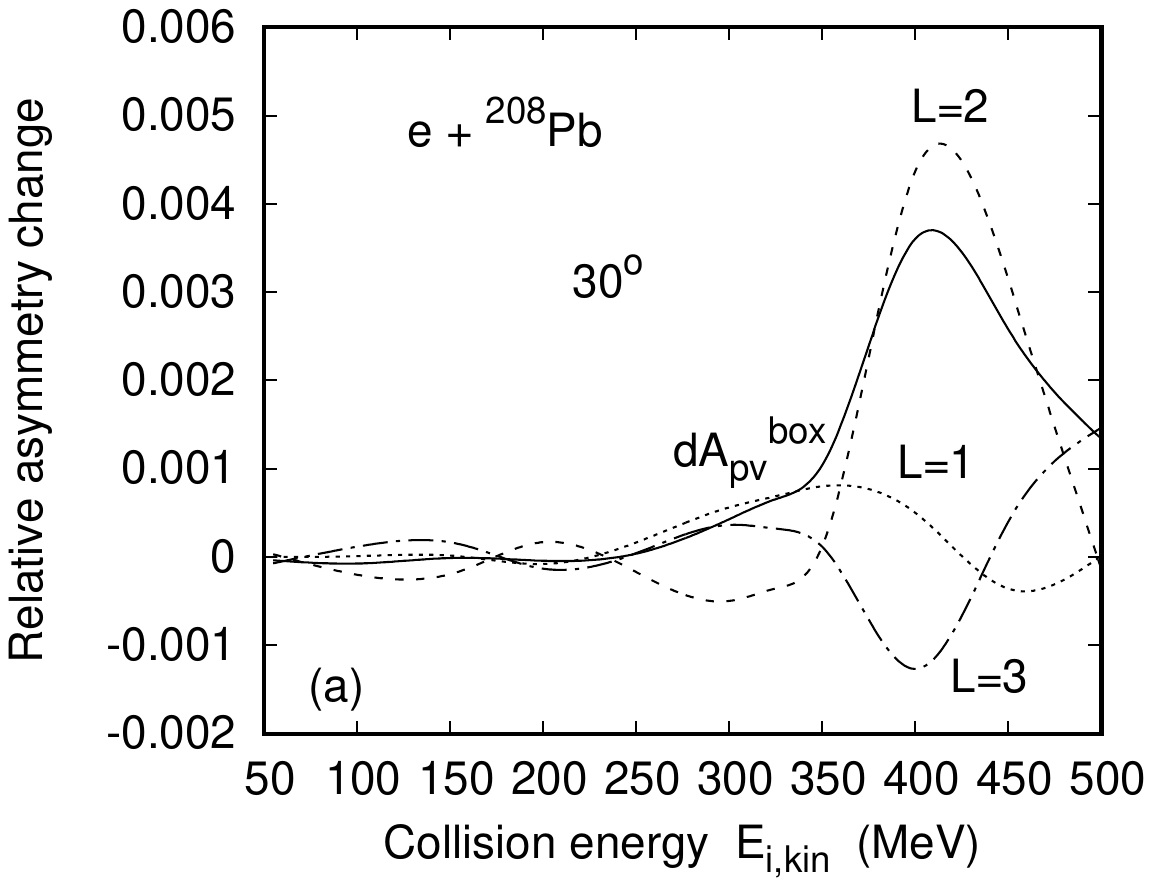}
\vspace{-1.5cm}
\vspace{-0.5cm}
\includegraphics[width=11cm]{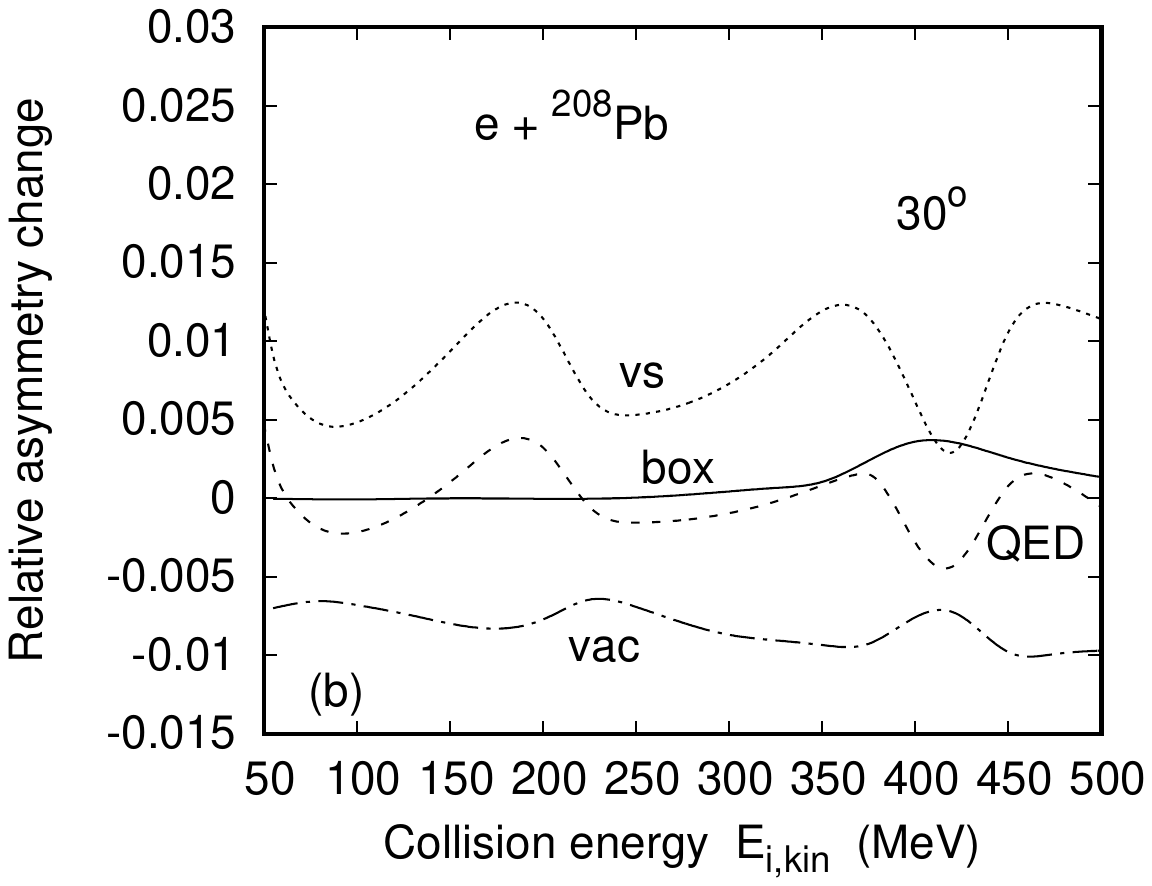}
\caption
{
Relative spin asymmetry change $dA_{\rm pv}$ in $e + ^{208}$Pb collisions at $\theta=30^\circ$ as function of collision energy.
Shown in (a) are the dispersive changes $dA_{\rm pv}^{\rm box}$ when summed over the dipole states $(\cdots\cdots)$,
over the quadrupole states $(----)$, over the octupole states $(-\cdot -\cdot -)$
and over all states with $L\leq 3$ ($dA_{\rm pv}^{\rm box}$, -------).
Shown in (b) are the QED changes by vacuum polarization $(-\cdot -\cdot -)$, by the vertex plus self-energy correction $(\cdots\cdots)$ and by both QED effects
$(dA_{\rm pv}^{\rm QED},\;----)$. Included is $dA_{\rm pv}^{\rm box}$ (----------) from (a).
\label{fig14}}
\end{figure}

The energy distribution of the relative spin asymmetry change is provided in Fig.~\ref{fig14} for a forward scattering angle.
The dispersive corrections (Fig.~\ref{fig14}a) are well below 0.1\% at energies below 300 MeV. However, in contrast to $^{12}$C at this angle (Fig.~\ref{fig7}), they increase with energy, showing a maximum of about 0.4\% in the second diffraction minimum, basically produced by the two isoscalar $2^+$ excitations. 
The QED corrections (Fig.~\ref{fig14}b) are larger, but still at most 0.5\% at this angle, irrespective of energy.

For the high-energy geometry, 953 MeV at $4.7^\circ$, our results are $A_{\rm pv}=5.896 \times 10^{-7},\; dA_{\rm pv}^{\rm QED}  \approx 8.79 \times 10^{-4}$ and $dA_{\rm pv}^{\rm box} =-8.52 \times 10^{-5}$,
confirming that dispersion as well as the QED effects are  negligible at GeV energies for small angles.

\section{Conclusion}
\label{conclusion}
We have estimated the radiative corrections to the parity-violating spin asymmetry for electron scattering from $^{12}$C and $^{208}$Pb nuclei by treating the QED effects nonperturbatively and 
by estimating the dispersive corrections from low-lying excited states by evaluating the $\gamma Z$ box diagrams in second-order Born approximation.
We have found that at small scattering angles and low  collision energies, the dispersive changes of $A_{\rm pv}$ are of little importance.

At backward angles 
and higher energies dispersion catches up for $^{12}$C, becoming equally important as the QED effects.
For $^{208}$Pb, on the other hand, the QED corrections are always the dominating ones.
For carbon, when the diffraction effects do not obscure the energy dependence of the spin asymmetry, we have found that dispersion decreases with energy beyond 150 MeV at forward angles.
This is in concord with the expectation that the contribution from the low-lying excited states is suppressed according to $1/E_i^2$ for energies well above the pion production  threshold \cite{Go24}.
On the other hand, the QED corrections are also small and nearly independent of collision energy, being usually below 1\%.
For the  test study at 953 MeV and $4.7^\circ$, the geometry selected in \cite{Ad21},  it is found that for $^{12}$C, the two radiative effects add to each other, however still affecting $A_{\rm pv}$ by less than one percent. For $^{208}$Pb they are of opposite sign and the result is one order of magnitude smaller than for $^{12}$C.

For the design of precision experiments it is important to choose small scattering angles, well below the onset of the diffraction structures.
One reason is that the radiative corrections, calculated within a given nuclear model, are tiny at small angles but increase with angle,
the more so the higher the collision energy.
Moreover, also the dependence of the ground-state charge density and of the 
parity-violating asymmetry on the underlying nuclear models is quite strong for the larger angles.
Also, whenever dispersion is important, the model-dependence of the transition densities cannot be neglected. 

\begin{acknowledgements}
It is a pleasure to thank A.V.Afanasev, M.Gorshteyn and C.J.Horowitz for many enlightening discussions. X.R.M. acknowledges support by MICIU/AEI/10.13039/501100011033 and by FEDER UE through grants PID2023-147112NB-C22; and through the “Unit of Excellence Maria de Maeztu 2025–2028” award to the Institute of Cosmos Sciences, grant CEX2024-001451-M. Additional support is provided by the Generalitat de Catalunya (AGAUR) through grant 2021SGR01095.
\end{acknowledgements}

\appendix

\section{Correspondence between electromagnetic and weak scattering}
\renewcommand{\theequation}{\Alph{section}.\arabic{equation}}

\vspace{0.2cm}
 In this Appendix we give some details on the correspondence between electromagnetic (two-photon) scattering and weak (photon-$Z$ boson) scattering.

\vspace{0.2cm}
For the weak interaction, the photon propagator is replaced by the $Z$-boson propagator, 
\begin{equation}\label{2.10}
\frac{g^{\mu \nu}}{q^2+i\epsilon} \,\mapsto\; \frac{g^{\mu \nu} -q^\mu q^\nu/(M_z^2c^2)}{q^2-M_z^2c^2+i\epsilon}\;\approx \frac{g^{\mu \nu}}{-M_z^2c^2},\;\mu ,\nu=0\mbox{-}3,
\end{equation}
where $g^{\mu\nu}$ is the metric tensor, $q$ the momentum of the photon, respectively $Z$ boson, and the $q$-dependence on the rhs is eliminated
due to the large mass $M_z\gg q/c$ of the $Z$ boson.

Moreover, the electron-$\gamma$ vertex is substituted by the electron-$Z$ vertex, relying on the Standard Model (and noting the different definition of $e$ in \cite{Th13}),
\begin{equation}\label{2.11}
\frac{i \sqrt{4\pi} e}{\sqrt{c}}\,[\bar{u}_f\, \gamma_\mu \,u_i] \mapsto -\;\frac{i}{4}\,g_z\,[\bar{u}_f \,\gamma_\mu\,(4\sin^2\theta_{\rm w} -1 +\gamma_5)\,u_i],
\end{equation}
where $u_f,u_i$ are plane-wave 4-spinors, $\gamma_\mu$ are Dirac matrices and $\gamma_5={0 \;I \choose I\;0}$ is the $4\times 4$ chirality matrix.
We have introduced the weak coupling constant $g_z$ of the $Z$ boson,
\begin{equation}\label{2.12}
g_z\,=\,\frac{g_{\rm w}}{\cos \theta_{\rm w}},\quad g_{\rm w}\,=\,\sqrt{\frac{8M_{\rm w}^2c^4G_F}{\sqrt{2}}},
\end{equation}
where $g_{\rm w}$ is the coupling constant of the $W$ boson, $\theta_{\rm w}$ is the Weinberg mixing angle ($\sin^2 \theta_{\rm w}=0.23857$ \cite{Co22})  and $M_{\rm w}c^2$ is the rest energy of the $W$ boson.

Likewise, the $\gamma$-nucleus vertex is replaced by the $Z$-nucleus vertex,
\begin{equation}\label{2.13}
-\,\frac{i \sqrt{4\pi} e}{\sqrt{c}}\;[\bar{\phi}_n\,\gamma_0\,J_{\rm em}^\mu\,\phi_n]\,\mapsto -\,\frac{i}{4}\,g_z\,[\bar{\phi}_n\,\gamma_0\,J_{\rm w}^\mu\,\phi_n],
\end{equation}
where $J_{\rm em}$ is the electromagnetic, $J_{\rm w}$ the weak transition current and $\phi_n$ a nuclear state.

Proton ($\phi_P$) as well as neutron ($\phi_N$) part of the nuclear wavefunction $\phi_n$ contribute to the matrix element of $J_{\rm w}^\mu$.
For a single proton $|P\rangle$ or neutron $|N\rangle$,
the isospin rotation model
provides the correspondence between $J_{\rm w}^\mu$ and the electromagnetic currrent,
 if the axial (magnetic) contribution to $J_{\rm w}^\mu$ is disregarded
 \cite{Go11,HP12},
$$\langle P|\,J_{\rm w}^\mu\,|P\rangle \,=\,(1-4\sin^2 \theta_{\rm w})\,\langle P|\,J_{\rm em}^\mu\,|P\rangle\,-\,\langle N|\,J_{\rm em}^\mu\,|N\rangle$$
\begin{equation}\label{2.14}
\langle N|\,J_{\rm w}^\mu\,|N\rangle \,=\,(1-4\sin^2\theta_{\rm w})\,\langle N|\,J_{\rm em}^\mu\,|N\rangle \,-\,\langle P|\,J_{\rm em}^\mu\,|P\rangle ,
\end{equation}
Generalizing (\ref{2.14}) by means of $\langle \phi_P | J^\mu|\phi_P\rangle = Z\,\langle P|J^\mu|P\rangle$ and $\langle \phi_N|J^\mu|\phi_N\rangle = N\,\langle N|J^\mu|N\rangle$, 
where the prefactors $Z$ and $N$ are the numbers of protons and neutrons in the nucleus, one obtains
$$\langle \phi_n|J_{\rm w}^\mu|\phi_n\rangle \,=\,\langle \phi_P | J_{\rm w}^\mu|\phi_P\rangle \,+\, \langle \phi_N|J_{\rm w}^\mu|\phi_N\rangle$$
\begin{equation}\label{2.15}
=\;(1-4\sin^2\theta_{\rm w}-\frac{N}{Z})\,\langle \phi_P|J_{\rm em}^\mu|\phi_P\rangle
\end{equation}
$$ +\;(1 -4 \sin^2\theta_{\rm w}-\frac{Z}{N})\,\langle \phi_N|J_{\rm em}^\mu|\phi_N\rangle.$$
In the special case of an $N=Z$ nucleus (like $^{12}$C), the $Z$-nucleus vertex simplifies to
$$-\,\frac{i}{4}\,g_z [\bar{\phi}_n \gamma_0 J_{\rm w}^\mu \phi_n]\,$$
\begin{equation}\label{2.16}
=\;i\,g_z \sin^2\theta_{\rm w}\left[ \langle \phi_P|J_{\rm em}^\mu|\phi_P\rangle +\langle \phi_N|J_{\rm em}^\mu|\phi_N\rangle \right].
\end{equation}


\section{Evaluation of the transition matrix element $M_{fi}^{Z\gamma}$ for diagram (a) of Fig.1.}

\vspace{0.2cm}
Substituting in the electromagnetic formalism \cite{Jaku22} the $Z$-boson propagator and vertices, one obtains
$$M_{fi}^{Z \gamma}(L,\omega_L,\pm)\,=\,-\;\frac{1}{(2\pi)^3}\;\frac{e^2}{c}\;\left(\frac{g_z}{4}\right)^2 $$
\begin{equation}\label{2.17} 
\times \;\int d\bfp\;\frac{1}{(-M_z^2 c^2)}\;\frac{1}{q_1^2+i\epsilon}\sum_{\mu \nu} t_{\mu \nu}^{Z\gamma}(\pm)\sum_M (T_n^{Z\gamma})^{\mu \nu},
\end{equation}
where $q_1=k_i-p$ and the sum runs over  the magnetic quantum number $M$ of the nuclear state with angular momentum $L$.

When helicity is conserved (corresponding to electron mass $m_e=0$), 
the electronic transition matrix element $t_{\mu \nu}^{Z \gamma}$ reads
$$t_{\mu \nu}^{Z\gamma}(\pm)\,=\,c\left[ \psi_{f,\pm}^{(0)+}\,(4\sin^2 \theta_{\rm w} -1+\gamma_5)\,\gamma_0 \gamma_\mu\right.$$
\begin{equation}\label{2.18}
\times\;\left. \frac{E_p+c\bfalpha \bfp +\gamma_0 c^2}{E_p^2-\bfp^2c^2 -c^4+i\epsilon}\, \gamma_0 \gamma_\nu \;\psi_{i,\pm}^{(0)}\right],
\end{equation}
where $E_p = E_i-\omega_L-\sqrt{\bfq_1^2 c^2 +M_T^2 c^4} + M_Tc^2$ is the energy of the electron in its intermediate state, 
with $M_T$ the target mass.
In the present calculations the electron mass is retained, allowing for spin-flip (where $\psi_{f,\pm}^{(0)}$ has to be replaced by  $\psi_{f,\mp}^{(0)}$).
However, it only plays any role at the lowest energies or backmost angles considered 
(for example, in 55 MeV $e+ ^{12}$C collisions  for  $\theta = 175^\circ$, the spin-flip contribution to the relative spin asymmetry change is actually 3\%).

The nuclear matrix element $(T_n^{Z\gamma})^{\mu \nu}$ for the transient transition from the ground state $|0\rangle$ to an excited state $|n\rangle$ which is
characterized by the quantum numbers $L$ and $M$,  is given by
$$(T_n^{Z\gamma})^{\mu \nu}\,=\,\langle 0|J_{\rm w}^\mu(\bfq_2)|n\rangle \; \langle n|J_P^\nu(\bfq_1)|0\rangle,$$
\begin{equation}\label{2.19}
J_{\rm w}^\mu(\bfq)=(1-4\sin^2 \theta_{\rm w}-\frac{N}{Z})J_P^\mu(\bfq)
\end{equation}
$$+\,(1-4\sin^2 \theta_{\rm w}-\frac{Z}{N})\,J_N^\mu(\bfq),$$
where $J_{\rm w}^\mu$ is the weak transition density, constructed in analogy to (\ref{2.15}) from the superposition of the proton $(J_P^\mu)$ and neutron ($J_N^\mu)$ 
parts of the charge ($\mu=0)$ and magnetic $(\mu=1,2,3)$ transition densities of the target nucleus.
Its matrix elements can be expressed
in terms of weak transition form factors \cite{Jaku22},
\begin{equation}\label{2.20}
\langle n|J_{\rm w}^\mu(\bfq)|0\rangle 
\end{equation}
$$=\left\{ \begin{array}{ll}
4\pi i^L F_L^{\rm c,w}(|\bfq|)\,Y_{LM}^\ast (\hat{\bfq}),& \mu=0\\
-4\pi i \sum_{\lambda=L\pm 1} i^\lambda F_{L\lambda}^{\rm te,w}(|\bfq|)(\bfY_{L\lambda}^{M\ast}(\hat{\bfq}))^m, & \mu=m
\end{array}\right.$$
with $m=1,2,3$ and $Y_{LM}$, respectively $\bfY_{L\lambda}^M$, a spherical harmonic function and a vector spherical harmonics.
The weak form factors $F_L^{\rm c,w}$ and $F_{L,L\pm 1}^{\rm te,w}$ are the Fourier transforms of the weak transition densities.
The photon-induced matrix elements $\langle n|J_P^\nu|0\rangle$ are also obtained from (\ref{2.20}) by replacing, respectively,  the weak form factors with the charge ($F_L^c)$ and the transverse electric $(F_{L\lambda}^{te})$ form factors.

For the further evaluation of $M_{fi}^{Z\gamma}$ the integrand in (\ref{2.17}) is written in terms of the photon propagator in Feynman gauge,
\begin{equation}\label{2.22a}
I(q)\, \equiv\,\frac{1}{q_1^2+i\epsilon} \;(T_n^{Z\gamma})^{\mu \nu}\,=\,\sum_{\varrho=0}^3 g^{\mu\varrho}
 \sum_{\tau=0}^3 \frac{g^{\nu \tau}}{q_1^2+i\epsilon}\;(T_n^{Z\gamma})_{\varrho \tau}.
\end{equation}
Upon decomposing the photon propagator into Coulombic and transverse terms
 \cite{Go61},
\begin{equation}\label{2.21}
\frac{g^{\nu\tau}}{q^2+i\epsilon}=-\,\frac{1}{\bfq^2}\,\delta_{\nu 0}\delta_{\tau 0}\,-\,\frac{\delta_{m k}-\hat{q}^m \hat{q}^k}{q^2+i\epsilon}\,\delta_{\nu m}\delta_{\tau k},\;m,k=1\mbox{-}3,
\end{equation}
where  $\hat{q}^m$ is the $m$-th component of the unit vector in direction of $\bfq$,
 one obtains
\begin{equation}\label{2.22}
\sum_{\mu \nu=0}^3 t_{\mu\nu}^{Z\gamma}\,I(q)\,=\;-\,\frac{1}{\bfq_1^{2}}\sum_{\mu=0}^3 (t^{Z\gamma})^\mu_{\;\,0}\,(T_{n}^{Z\gamma})_{\mu 0}\,
 -\,\frac{1}{q_1^2+i\epsilon}
\end{equation}
$$\times\; \left[ \sum_{m=1}^3 \sum_{\mu=0}^3 (t^{Z\gamma})^\mu_{\;m} \left( (T_{n}^{Z\gamma})_{\mu m} -\hat{q}_1^m\sum_{k=1}^3 \hat{q}_1^k (T_n^{Z\gamma})_{\mu k}\right) \right].$$
Corresponding to the two-term expression (\ref{2.22}) the transition matrix element is split into two parts, $M_{fi}^{Z\gamma}=M_1+M_2.$
Let us first concentrate on the further evaluation of $M_1$.
The singularity from the factor $1/\bfq_1^2$  in the first term of (\ref{2.22}) is omitted by changing the integration variable over the electronic intermediate states from $\bfp$ to $\bfp'=\bfp-\bfk_i=-\bfq_1.$
Upon insertion of 
(\ref{2.18}) and (\ref{2.22}) into (\ref{2.17}) one obtains (with $|\bfp'| \equiv p'$)
$$M_1\,=\,-\,\frac{g_z^2}{128 \pi^3 M_Z^2c^2} \int_0^\infty dp' \int d\Omega_{p'}$$
\begin{equation}\label{2.23}
\times\;\frac{1}{E_p^2-(\bfp'+\bfk_i)^2c^2 -c^4+i\epsilon}\,\sum_{\mu=0}^3 S_1(\bfp'),
\end{equation}
where $S_1(\bfp')$ is a well-behaved function,
$$S_1(\bfp')\,=\,\left[ \psi_{f,\pm}^{(0)+} (4 \sin^2 \theta_{\rm w} -1 +\gamma_5)\,\gamma_0\gamma^\mu\right.$$
\begin{equation}\label{2.24}
\left. \times\;(E_p+c\bfalpha (\bfp'+\bfk_i)+\gamma_0c^2)\,\psi_{i,\pm}^{(0)}\right]
\;\sum_M (T_n^{Z\gamma})_{\mu 0}.
\end{equation}

The structure of $M_2$ is analogous to (\ref{2.23}),
except for the additional factor $p^{'2}/[(E_i-E_p)^2/c^2-p'^{2} + i\epsilon]$,
which leads to a second singularity near $p'=\omega_L/c$.
This means that the evaluation of $M_{fi}^{Z\gamma}$ is done with the same techniques that are applied to the transition amplitudes
$M_{fi}^c$ and $M_{fi}^{te1}$ of the two-photon box diagram.
The basic difference is that the innermost integral (over the azimuthal angle of $\bfp'$) does not contain any singularity. 
This speeds up convergence and avoids the logarithmic singularity which occurs in the cross section for the
two-photon box diagram.
Nevertheless, the absence of $1/(q_2^2+i\epsilon)$
with $q_2=p-k_f$ in the integrand leads to a poorer convergence with respect to the upper integration limit $p_{\rm max}$ of the radial integral, which therefore relies solely on the vanishing of the transition form factors at large momentum transfers. 

Diagram (b) of Fig.1 is treated in the same way, but changing variables from $\bfp$ to $\bfp'=\bfp-\bfk_f=\bfq_2$. 
Therefore, also here, the same techniques can be used as for the evaluation of the transition amplitudes $M_{fi}^{te2}$ and $M_{fi}^{te3}$ of the two-photon box diagram \cite{Jaku22}.

\vspace{1cm}


\end{document}